\documentclass[10pt, a4paper]{article}
\usepackage[utf8]{inputenc}
\usepackage{authblk}
\usepackage{cite}
\usepackage{fullpage}
\usepackage{amsmath}
\usepackage{graphicx}
\usepackage{tabularx}
\usepackage{url}
\usepackage{titlesec}
\usepackage{multirow}
\usepackage{setspace}
\title{Frequency and frequency modulation share the same predictive encoding mechanisms in human auditory cortex}
\author[1]{Jasmin Stein}
\author[1]{Katharina von Kriegstein}
\author[1]{Alejandro Tabas}
\affil[1]{Faculty of Psychology, Technische Universit\"{a}t Dresden, Dresden, Germany}
\date{}

\date{}
\begin{document}

\maketitle

\begin{abstract}
    Expectations can substantially influence perception. Predictive coding is a theory of sensory processing that aims to explain the neural mechanisms underlying the effect of expectations in sensory processing. Its main assumption is that sensory neurons encode prediction error with respect to expected sensory input. Neural populations encoding prediction error have been previously reported in the human auditory cortex (AC); however, most studies focused on the encoding of pure tones and induced expectations by stimulus repetition, potentially confounding prediction error with effects of neural habituation. Here, we systematically studied prediction error to pure tones and fast frequency modulated (FM) sweeps across different auditory cortical fields in humans. We conducted two fMRI experiments, each using one type of stimulus. We measured BOLD responses across the bilateral auditory cortical fields Te1.0, Te1.1, Te1.2, and Te3 while participants listened to sequences of sounds. We induced subjective expectations on the incoming sounds independently of stimulus repetition using abstract rules. Our results indicate that pure tones and FM-sweeps are encoded as prediction error with respect to the participants' expectations across auditory cortical fields. The topographical distribution of neural populations encoding prediction error to pure tones and FM-sweeps was highly correlated in left Te1.1 and Te1.2, and in bilateral Te3, suggesting that predictive coding is the general encoding mechanism in AC.
\end{abstract}

\section*{Introduction}

    %\paragraph{Expectations shape perception}
    Subjective expectations influence our perception of the world \cite{de2018expectations}. They facilitate perceiving noisy \cite{leonard2016perceptual, chalk2010rapidly} or ambiguous \cite{chambers2017prior, sterzer2008believing} sensory input, and bias perception when inputs are overly expected \cite{lange2009brain}. Understanding the mechanisms integrating expectations with sensory input is an essential prerequisite for understanding perception. The predictive coding framework is a theory of sensory processing aiming to explain these mechanisms. Its main tenet is that sensory neurons encode prediction error with respect to an internal generative model of the sensory world \cite{rao1999predictive, mumford1992computational, friston2003learning, spratling2017review}.

    %\paragraph{PC in AC}
    Neurons \cite{nieto2016topographic, rubin2016representation, parras2017neurons, perez2021deviance} and neural populations \cite{cacciaglia2019auditory, zvyagintsev2020auditory} of the auditory cortex (AC) encode pure tones as prediction error. Prediction error is typically elicited using oddball paradigms, where predictable repetitions of a standard sound are rarely interrupted by a deviant. Individual neurons in the AC show reduced responses to repeated standards and recovered responses to deviants, a phenomenon that is called stimulus-specific adaptation (SSA) and typically interpreted as prediction error \cite{ulanovsky2003processing}. 
        
    However, whether SSA truly represents prediction error is unclear: its phenomenology can be explained by habituation to local stimulus statistics \cite{eytan2003selective, mill2011neurocomputational, wang2014stimulus} (see~\cite{tabas2021adjudicating} for review, and~ \cite{carbajal2018neuronal} for a different perspective). One way to disambiguate habituation and prediction error is to manipulate participants' subjective stimulus expectations orthogonally to local stimulus statistics \cite{tabas2020abstract, tabas2021adjudicating}. Prediction error to subjective expectations has been studied measuring the mismatch negativity (MMN) \cite{costa2011interactions, todorovic2011prior, cornella2012detection,todorovic2012repetition, lecaignard2015implicit, durschmid2016hierarchy, phillips2016convergent}, partially generated in the frontal cortex \cite{shalgi2007direct, deouell2007frontal}. Only two previous studies have measured prediction error to subjective expectations in the human AC using fMRI \cite{berlot2018mapping, cacciaglia2019auditory}.
    
    Moreover, these studies did not consider whether prediction error is encoded in all fields of the human AC. The auditory system consists of a primary and secondary subdivision \cite{lee2011classification}. Primary areas show narrow tuning curves; secondary areas are tuned more widely and support multi-sensory integration \cite{hu2003functional}. In rodents, SSA is stronger in secondary subdivisions of AC \cite{nieto2016topographic, parras2017neurons} and the subcortical auditory pathway \cite{antunes2010stimulus, antunes2011effect, duque2012topographic, duque2014modulation, ayala2015differences, parras2017neurons}. Whether the human AC shows the same organization is unknown.
    
    %\paragraph{Prediction error in AC}
    Additionally, most previous research on predictive coding in the AC has focused on pure tones. However, naturalistic auditory scenes entail different components. One important type of stimulus is frequency modulated (FM) sweeps, the basic information-bearing unit of animal \cite{Suga2012} and human \cite{divenyi2009perception} vocalization. In human speech, FM-sweeps constitute formant transitions, the main components of consonants preceding a vowel \cite{liberman1967perception}, which are critical for phoneme identification. While pure tones are first encoded in the basilar membrane \cite{malmierca2010structural}, FM selectivity is present only in the auditory midbrain, thalamus, and cortex \cite{geis2013intracellular, lui2003frequency, issa2017multiscale, hall2000modulation, hart2003amplitude, paltoglou2011mapping, altmann2014representation}. Given this fundamental difference, it is unclear whether the neural populations encoding prediction error to pure tones are also in charge of processing FM. Research on prediction error encoding of FM-sweeps in human cerebral cortex yielded mixed results: Some reported enhanced neural responses to repeated FM-stimuli, concluding that predictive encoding of FM differs from pure tones \cite{heinemann2010repetition, altmann2011repetition, heinemann2011auditory, okamoto2017modulation}. Others reported an MMN to deviating FM-stimuli, suggesting that FM is also encoded as prediction error \cite{cornella2013regularity, kung2020minimum, hsieh2021interaction}. However, E/MEG studies lack the spatial resolution to precisely locate neural populations generating these responses. The only fMRI study that has investigated prediction error to FM focused on subcortical areas \cite{tabas2021fast}. It is still unclear whether FM is encoded as prediction error in the AC and, if so, whether the same neural populations encode prediction error to pure tones and FM-sweeps.
    
    %\paragraph{Research goals}
    Here, we investigated whether and how prediction error to pure tones and FM-sweeps is encoded in human AC. We addressed three questions: First, whether the prevalence and topographic distribution of SSA to pure tones generalize to FM-sweeps across auditory cortical fields. Second, whether SSA to pure tones and FM-sweeps in each field reflects prediction error with respect to subjective expectations. Third, whether the same neural populations encode prediction error to pure tones and to FM-sweeps.

\section*{Methods}

    \subsection*{Data acquisition and experimental design}
    
        We analyzed fMRI-data from two previous studies that focused on subcortical auditory prediction error encoding of pure tones \cite{tabas2020abstract} and FM-sweeps \cite{tabas2021fast}. Both experiments used the same auditory oddball paradigm with different stimulus types. They were both acquired at different scanning sites and with different participants.

        \subsubsection*{Participants}

            All participants were neurotypical normal-hearing German native speakers (see~\cite{tabas2020abstract} and~\cite{tabas2021fast} for further details on inclusion criteria). Nineteen participants (12 female) between the ages of 24 and 34 (average 26.6) participated in the pure tone study; eighteen participants (12 female) between the age of 19 and 31 (average 24.6) participated in the FM-sweep study.
        
        \subsubsection*{Stimuli}
        
\begin{figure}[h!]
            \centering
            \includegraphics[scale=1]{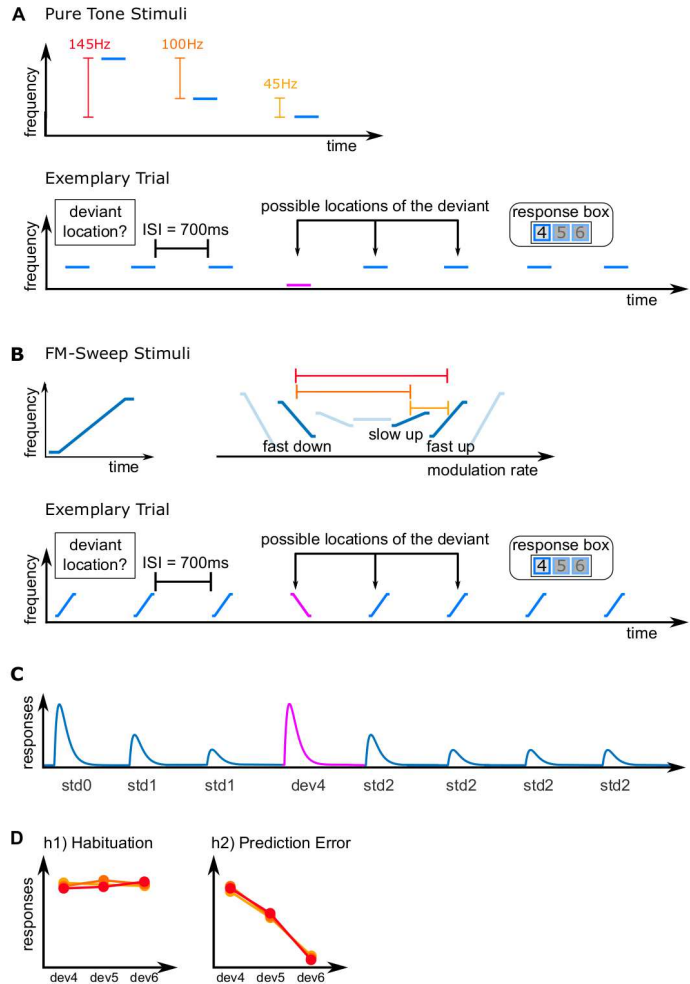}
            \caption[Experimental Design]{\textbf{Experimental design and hypotheses.} A: Pure tone oddball experiment; top: three pure tones with different frequencies were used to build six standard-deviant combinations with three absolute frequency spacings $\Delta = |f_{std} - f_{dev}|$; bottom: an example of a trial consisting of seven repetitions of a standard and one frequency deviant. B: FM-sweep experiment; top: FM-sweeps were used to build standard-deviant combinations, characterized by the absolute difference between the frequency spans of standard and deviant sweeps ($\Delta = |\Delta f_{dev} - \Delta f_{std}|$); bottom: an example of a trial consisting of seven repetitions of a standard and one deviant sweep. C: expected neural response to the exemplary trials shown in A and B (note the recovered response to the deviant as the main characteristic of SSA). D: two possible anticipated outcomes of the experiment; in h1 or habituation, it was assumed that high-level subjective expectations do not affect deviant responses; in h2 or prediction error, we expected sounds to be scaled by stimulus predictability and thus represent prediction error responses with respect to subjective expectations.}
            \label{fig:design}
            \end{figure}

            In the pure tone experiment, there were three pure tone stimuli of 50\,ms duration (including 5\,ms onset/offset ramps) and frequencies of 1455\,Hz, 1500\,Hz, and 1600\,Hz. The tones were combined into six pairings of standard and deviant tones. Each of the resulting oddball sequences was consequently characterized by one of three possible absolute frequency differences characterizing the distance between standards and deviants ($\Delta = |f_{std} - f_{dev}|$ either 145\,Hz, 100\,Hz, or 45\,Hz. See Figure \ref{fig:design}A for a visualization of sound stimuli and sequences.
            
            In the FM-sweep experiment, there were three sinusoidal FM-sweeps with a duration of 50\,ms (5\,ms onset/offset ramps) and starting frequencies of 1000\,Hz, 1070\,Hz, or 1280\,Hz. The FM-sweeps ended at either 1080\,Hz, 1170\,Hz, or 1200\,Hz, respectively. The tones were again combined into six pairings of standard and deviant tones. An FM-sweep could deviate from the standard in its FM direction, FM rate, or both. Since all sweeps had the same duration, the defining property of the FM-sweeps was their frequency span $\Delta f$---the difference between starting and ending frequencies. Each of the sequences was consequently characterized by one of three possible absolute frequency span differences characterizing the distance between standards and deviants ($\Delta = |\Delta f_{dev} - \Delta f_{std}|$). See Figure \ref{fig:design}B for an exemplary illustration of the FM-sweep stimuli and sequences \cite{tabas2021fast}.
            \clearpage
            
        \subsubsection*{Experimental Design}
        
            The design was a variation of the oddball paradigm where abstract rules were used to manipulate participants’ high-level expectations on the upcoming stimuli independent of the local statistical regularity of the presented sound sequences. Specifically, participants listened to sequences of eight sounds: seven repeated standards and one deviant that could occur in positions four, five, or six. The stimuli were separated by a 700\,ms inter-stimulus interval. The inter-trial interval was jittered (minimum: 1500\,ms, maximum: 11\,s). All sound combinations were used equally often across runs assuring that all sound types were used as standards and deviants the same number of times.

            Participants were told explicitly that all sequences would contain a deviant, and that the deviant would occur in one of the three aforementioned positions. The participants were instructed to respond to the presentation of the deviant via button press as fast and as accurately as possible. Deviants are equally likely to be placed in each of the three positions at the beginning of each trial. Thus, the probability of a deviant in position four after hearing three standards is 1/3. However, if the deviant was not in position 4, since deviants occur once in each sequence, the probability of hearing a deviant in position five after hearing four standards is 1/2. If the deviant was neither in position 4 nor 5, the probability of a deviant in position six after hearing five standards is 1 \cite{tabas2020abstract, tabas2021fast}.

            The pure tone experiment comprised four runs that were completed by all participants. The FM-sweep data was collected in three sessions with three runs each; most participants completed 9 runs, one participant only completed eight due to technical reasons. In both experiments, a run contained six blocks of ten trials. Deviant positions were pseudo-randomized so that they all occurred 20 times in each run. The runs lasted for around ten minutes and were separated by a one-minute break. Practice trials were presented at the beginning of the first run to ensure task understanding. Interspersed null events were used to optimize the fit of the GLMs \cite{friston1999stochastic}. Further details can be found in~\cite{tabas2020abstract} (pure tones) and~\cite{tabas2021fast} (FM-sweeps).
    
        \subsubsection*{FMRI data acquisition}
        
            FMRI data were collected using EPI sequences and partial FoVs. Magnetic field strength and image resolution differed between data sets. Data from the pure tone experiment was collected using a Siemens Magnetom 7\,Tesla scanner (Siemens Healthineers, Erlangen, Germany) with an 8-channel head coil and a voxel size of 1.5\,mm isotropic; data from the FM-sweep experiment was collected using a Siemens Trio 3\,Tesla scanner (Siemens Healthineers, Erlangen, Germany) with a 32-channel head coil and a voxel size of 1.75\,mm isotropic. Interleaved slice acquisition was used in both data sets.
        
            Pure tone data were collected using the following scanning parameters: $TR = 1600$\,ms, $TE = 19$\,ms, flip angle $65^{\circ}$, GRAPPA with acceleration factor 2, 33\% phase oversampling, matrix size $88\times88$, phase partial Fourier 6/8, $FoV = 132$\,mm$\times$132\,mm (30 slices). We also acquired three whole-head volumes with 80 slices to aid coregistration. FM-sweep data were collected using the following scanning parameters: $TR = 1900$\,ms, $TE = 42.2$\,ms, flip angle $66^{\circ}$, matrix size $88\times88$, $FoV = 154$\,mm$\times$154\,mm (24 slices).  We also acquired one whole head volumes with 84 slices to aid the coregistration process.
            
            Structural images for the pure tone experiment were measured using an MP2RAGE T1 protocol (700\,mm isotropic resolution, $TE = 2.45$\,ms, TR = 5000$\,ms, TI1 = 900\,ms$, $TI2 = 2750$\,ms, flip angle $1 = 5^{\circ}$, flip angle $2 = 3^{\circ}$, $FoV = 224$\,mm$\times$224\,mm, GRAPPA acceleration factor 2). Structural images for the FM-sweep data were measured using an MPRAGE T1 protocol (1\,mm isotropic resolution, $TE = 1.95$\,ms, $TR = 1000$\,ms, flip angle $1 = 8^{\circ}$, $FoV = 256$\,mm$\times$256\,mm).
             
            Physiological data (heart rate and respiration in the pure tone experiment, heart rate in the FM-sweep experiment) were collected and processed for use as regressors of no-interest during model estimation for both sound modalities.

    \subsection*{Data preprocessing}
        
        \subsubsection*{Anatomical data}
    
            Data preprocessing and analysis was implemented using Nipype 1.1.2 \cite{gorgolewski2011nipype} and included functions from: the FMRIB Software Library, version 5 (FSL) \cite{jenkinson2012fsl}; Freesurfer, version 7 \cite{fischl2002van}; the Advanced Normalization Tools, version 2.2.0 (ANTS) \cite{avants2011reproducible}; and the Statistical Parametric Mapping toolbox (SPM) \cite{penny2011statistical}, version 12.
            
            All anatomical data was resampled to a resolution of 1\,mm isotropic. We computed the boundaries between gray and white matter using Freesurfer's \emph{recon-all}. These boundaries were later used for coregistration of the functional data to the participants' structural images. In the case of the pure tone experiment, we first computed a brain mask excluding voxels containing air, cerebrospinal fluid, scalp, and skull. This was necessary because MP2RAGE (but not MPRAGE) yields noisy signals outside the brain that interfere with the automatic processes of \emph{recon-all}. The mask was calculated using Freesurfer's \emph{BET} and SPM's \emph{Segment} and was applied using \emph{FSLmath}. Then, Freesurfer's \emph{recon-all} was used to obtain gray and white matter boundaries, and ANTs was used to calculate the coregistration matrix between the anatomical data and the MNI152 symmetric template.
    
        \subsubsection*{FMRI data}
        
            We used SPM's \emph{FieldMap Toolbox} to calculate distortions due to magnetic field inhomogeneity. Then, motion and distortion correction was performed on the functional data separately for each session (SPM \emph{Realign and Unwarp}). Nipype module's \emph{rapidart} was used to detect artifacts from the realigned functional data to serve as regressors of no-interest in our design matrix during GLM estimation. The resulting functional data were smoothed (SPM \emph{Smooth}) using a 2\,mm FWHM Gaussian kernel. 
            In the case of the pure tone data, the derivatives (i.e., log-evidences and beta maps) were registered to the anatomical space after fitting (see \emph{GLM Estimation} and \emph{Bayesian Model Comparison}). For FM-sweeps, the realigned functional data were registered to the anatomical space using Freesurfer's \textit{ApplyVolTransform} before model estimation to ensure all data was available in the same space during model fitting. 
            
            The transformation matrix between functional and structural data was computed using Freesurfer's \emph{BBRegister} using the white and gray matter boundaries computed as described above and the whole-brain EPI as an intermediate stage.
    
    \subsection*{Anatomical ROIs}
    
        Anatomical ROI were taken from Morosan et al. \cite{morosan2001human}. The ROIs of interest are the bilateral auditory cortical fields Te1.0, Te1.1, Te1.2, and Te3 \cite{morosan2001human}. Areas Te1.0, Te1.1, and Te1.2 are mostly located on Heschl's gyrus (Te1.1 most medial, Te1.2 most lateral). These areas were originally considered A1 \cite{morosan2001human}. Te1.0 and Te1.1 have been proposed to correspond mostly to BA 41. Te1.2 also overlaps with BA 42 \cite{moerel2014anatomical}. Comparing human and primate auditory fields, it was assumed that Te1.0 corresponds to the auditory core and Te1.1 and Te1.2 represent medial junction and lateral belt \cite{moerel2014anatomical}. Te3 lies on the lateral surface of the superior temporal gyrus, is an auditory association area, part of BA 22 and might correspond to parabelt areas in primates \cite{moerel2014anatomical}. However, functional differences between human auditory fields and their correspondence to primates' auditory fields are still unclear; e.g., \cite{moerel2014anatomical, besle2019human, gulban2020improving}.    
        
    \subsection*{Data analysis}
        
        \subsubsection*{GLM Estimation}
            
            First level analyses were performed using SPM's \emph{EstimateModel}. Statistical analysis at the participant- and group-level were conducted in MATLAB (The MathWorks Inc., Version 2020b) using custom code. 
            
            We estimated one GLM per participant. The model included six task regressors: $std0$ (the first standard in a sequence), $std1$ (standards before the deviant), $std2$ (standards after the deviant), $dev4$, $dev5$, and $dev6$ (deviants in positions four, five, and six). The first standard was modeled as a separate regressor to test for adaptation by comparing the estimates corresponding to $std0$ and $std1$/$std2$. $std1$ and $std2$ were parametrically modulated in a linear fashion according to their position relative to the deviant: values corresponding to $std1$ were assigned amplitudes from one to the total number of $std1$ in the sequence and $std2$ were assigned amplitudes from one to the total number of $std2$ in the sequence. This was done to account for a slight recovery of standard responses after the occurrence of a deviant. Thus, for example, in a sequence with a deviant in position four, $std1$ were assigned the amplitudes amp$_{1}$ = [1, 2] and $std2$ were assigned the amplitudes amp$_{2}$ = [1, 2, 3, 4]. Choosing increasing (i.e.,~$\text{amp} = 1, 2, \dots$) instead of decreasing ($\text{amp} = 4, 3, \dots$) amplitudes does not change the results since the parametric modulator is used as a regressor of no-interest. All amplitudes were z-standardized before model fitting. Note that SPM does not prime positive over negative regressors (i.e., the results are symmetric under linear transformations of the amplitudes). In addition, we added physiological data, artifact regressors, and realignment parameters to the design matrix as regressors of no interest. 
    
            Model estimations for the pure tone data were done using the smoothed data in the native space of the functional data of the individual participants. For FM-sweeps, we estimated the models in the space of the participants' anatomical scans. After model estimation, the spatial transformations calculated before were applied to the resulting statistical maps. The statistical maps of the pure tone data were first registered to the participants' anatomical scans using Freesurfer's \emph{ApplyVolTransform} and subsequently to the MNI152 symmetric template using ANTs' \emph{ApplyTransforms}. For FM-sweeps, statistical maps were registered directly to MNI space.
                
            The resulting beta estimates were $z$-standardized according to participant, experimental run, and ROI before the second level analyses to reduce variance specific to participants, runs, and ROIs.

        \subsubsection*{Identifying Voxels Showing SSA}
        
            To localize voxels showing SSA, we first identified voxels within the anatomical ROIs showing adaptation (reduced responses to repeated standards) and deviant detection (stronger response to deviants compared to standards). Adapting voxels were identified using the contrast $std0 > 0.5\,std1 + 0.5\,std2$ and deviant detecting voxels were identified using the contrast $dev4 > 0.5\,std1 + 0.5\,std2$. We only included $dev4$ in the latter contrast since this is the only deviant for which predictive coding and habituation make the same prediction. We tested both contrasts using right-tailed rank-sum tests. Before conducting the rank-sum tests, we averaged single-voxel beta estimates for each experimental condition across all experimental runs.

            We defined SSA regions for each stimulus type as the set of voxels showing significant adaptation and deviant detection: We computed voxel-wise $p$-values for SSA as the maximum of the uncorrected $p$-values for adaptation and deviant detection in each voxel; $p_{SSA} = max(p_{adaptation}, p_{deviant~detection})$. All voxels' $p$-values were subsequently controlled for the false discovery rate (FDR) using the Benjamini-Hochberg method \cite{benjamini1995controlling} and thresholded at $\alpha = 0.05$. Peak-level $p$-values were corrected for the FWE rate: we corrected for the number of voxels per ROI using Bonferroni-correction and for the total number of comparisons using the Holm-Bonferroni method \cite{holm1979simple}.

        \subsubsection*{Quantifying SSA Magnitude}
        
            To quantify SSA magnitude in each voxel of the anatomical ROIs, we computed the standardized voxel-wise index of SSA (iSSA) following the procedure described in previous research \cite{ulanovsky2003processing, tabas2020abstract}: We normalized the beta estimates for $dev4$, $std1$, and $std2$ to a range from zero to one, averaged these values in each voxel across participants and runs, and computed the index of SSA as $\text{iSSA} = (dev4 - 0.5\,std1 - 0.5\,std2) / (dev4 + 0.5\,std1 + 0.5\,std2)$. To test if our voxel-wise results were reproducible across sound modalities, we computed Pearson correlations between SSA magnitude for pure tones and FM-sweeps in each anatomical ROI.

        \subsubsection*{Classical Analysis}
       
            For both sound modalities, we conducted a classical statistical analysis to test for differences between responses to deviants in positions four, five, and six. To specifically investigate mechanisms driving SSA, we restricted this analysis to SSA clusters with significant ($p < 0.05$, FWE-corrected) peak-level $p$-values---the SSA ROIs.

            We tested the pairwise differences between responses to deviants in different positions ($dev4 > dev5$, $dev4 > dev6$, and $dev5 > dev6$) in each SSA ROI using one-sided Wilcoxon sign rank tests at the group-level. Before testing the contrasts, we averaged data corresponding to each experimental condition across runs and voxels within each participant and SSA ROI. In line with the idea of prediction error encoding, we expected deviant responses to be stronger when deviants are less expected. 
            
            We also measured the effect size of adaptation and deviant detection by testing the contrasts $std0 > std2$ and $dev4 > std2$. Note that these contrasts are not independent of the contrasts used for SSA voxel selection. However, we included them here to be able to quantify the size of both effects. Additionally, we included the comparison of $dev6$ and $std2$ using two-tailed Wilcoxon sign rank tests. We included this analysis to test whether responses to fully predictable deviants were comparable to standard responses. In line with predictive coding, we expected no statistically significant difference between responses to $dev6$ and $std2$. That is because both types of sound are fully predictable given our experimental design. All $p$-values were corrected for multiple comparisons using the Holm-Bonferroni method \cite{holm1979simple}.

        \subsubsection*{Correlational Analysis and Linear Mixed-Effects Model}
        
            To investigate the hypothesized negative relationship between deviant predictability and deviant responses further, we estimated a linear mixed-effects model in each SSA ROI for each data set at the group-level. For pure tones, the model included deviant predictability as a fixed effect and random intercepts and slopes for experimental runs and participants:
            
            \[
                \text{beta} \sim~1 + \text{predictability} + (1 + \text {predictability}|\text{run}) + (1 + \text{predictability}|\text{participant})
            \]
            
            For FM-sweeps, we used the same model, but added experimental session as an additional random effect. All $p$-values were Bonferroni-corrected for the total number of SSA ROIs. To test if the group-level results were replicated at the participant-level, we computed Spearman’s rank correlation between deviant predictability (1/3 for $dev4$, 1/2 for $dev5$, and 1 for $dev6$) and standardized beta estimates for $dev4$, $dev5$, and $dev6$ in each participant. Before computing the correlation coefficient $\rho$, beta estimates for the different deviant conditions in each voxel were averaged across experimental runs.
            
        \subsubsection*{Bayesian Model Comparison}
        
            We constructed two models, each representing one potential encoding mechanism driving SSA (see Figure \ref{fig:design}D. The models were defined using parametric amplitude modulation vectors that specified the predicted responses to all tones in each trial.
    
            \paragraph{h1) Habituation:} SSA is based on stimulus repetition. Responses to standards undergo habituation over time and recover slightly after the deviant. Deviant responses are fully recovered and do not differ between deviants of differential predictability.
                
                The h1 model (see Figure \ref{fig:design}D, left) was specified by assigning the amplitude 1 to \textit{std0} and the deviant of a sequence. Standards before the deviant were assigned the amplitudes $1/n$ and standards after the deviant were assigned the amplitudes $1/(n - 1)$, where $n$ is the position of the standard within the sequence (see Table \ref{tab:BMC} for the exact amplitudes). 
    
            \paragraph{h2) Prediction error:} Neural responses to sounds are scaled by stimulus predictability and thus represent prediction error responses. Subjective expectations drive deviant responses, which are stronger when stimuli are less expected.
    
                The h2 model (see Figure \ref{fig:design}D, right) used an amplitude of 0.5 for $std0$ and an amplitude of $p$ ($p$ = probability of stimulus occurrence) for the rest of the tones. For example, a sequence with a deviant in position five was assigned the amplitude vector amp$_{0}$ = [1/2, 1, 1, 2/3, 1/2, 1, 1, 1]. Thus, $std4$ was assigned a value of 2/3 since a standard in position four is expected with a probability of 2/3, and $dev5$ was assigned a value of 1/2 because deviants in position five are expected with a probability of 1/2 after hearing four standards. See Table \ref{tab:BMC} for all amplitudes of h2. Since model estimation using parametric modulation is symmetric with respect to linear transformations of the amplitudes, the above-described amplitudes are equivalent to assuming a decreasing response with increasing sound predictability.
                
                For each subject, we computed the log-evidence of the two models in each voxel of all anatomical ROIs using SPM's Bayesian Estimation functions in Nipype. Before model fitting, the amplitudes of all models were z-standardized according to experimental runs.
        
                For pure tones, all models were estimated using the smoothed functional data in their native space. The log-evidence maps were registered to the individual T1 scans and then to the MNI152 symmetric template using Freesurfer's \emph{ApplyVolTransform} and ANTs' \emph{ApplyTransforms}, respectively. We combined the log-evidence-maps of all participants and calculated posterior probability maps for each model using custom code by Tabas et al. (2020) \cite{tabas2020abstract} following the methodology described in \cite{rosa2010bayesian, stephan2009bayesian}.
                
                For FM-sweeps, all models were estimated using the smoothed functional data across sessions in the space of the participants' anatomical scans. The resulting log-evidence-map of each participant was registered to the MNI152 symmetric template using ANTs' \emph{ApplyTransforms} and posterior probability maps were calculated as described before.
                
                Then, we computed the Bayes factor K for both models in each voxel of our anatomical ROIs. To test if the spatial distribution of prediction error encoding is similar for pure tones and FM-sweeps, we computed Pearson correlations between the Bayes factor maps of both sound modalities in each anatomical ROI.
                
                \begin{table}[h!]
                    \centering
                    \begin{tabular}{cccccccccc}
                        \textbf{h1} & deviant position &  1  &  2  &  3  &        4     &       5      &     6    &  7  &  8  \\ 
                        \hline
                                    &                4 &  1  & 1/2 & 1/3 &        1     &      1/3     &    1/4   & 1/5 & 1/6 \\
                        a$_{0}$     &                5 &  1  & 1/2 & 1/3 &       1/4    &       1      &    1/4   & 1/5 & 1/6 \\
                                    &                6 &  1  & 1/2 & 1/3 &       1/4    &      1/5     &     1    & 1/5 & 1/6 \\
                        \hline
                        \\
                        \textbf{h2} & deviant position &  1  &  2  &  3  &        4     &       5      &     6    &  7  &  8  \\ 
                        \hline
                                    &                4 & 1/2 &  1  &  1  &      1/3     &       1      &     1    &  1  &  1  \\
                        a$_{0}$     &                5 & 1/2 &  1  &  1  &      2/3     &      1/2     &     1    &  1  &  1  \\
                                    &                6 & 1/2 &  1  &  1  &      2/3     &      1/2     &     1    &  1  &  1  \\
                        \hline
                        \\
                    \end{tabular}
                    \caption[Bayesian Model Comparison: Amplitudes]{\textbf{Amplitudes of the models used for Bayesian model comparison.} h1 (habituation): asymptotically decreasing responses to repeating standards, full responses to deviants, and a slight recovery after the deviant; h2 (prediction error): neural responses scaled by sound predictability, amplitude defined as the probability $P$ of finding the heard sound in each position of the sequence.}
                    \label{tab:BMC}
                \end{table}

\section*{Results}

    \subsection*{Topography of SSA to pure tones and FM-sweeps in AC}
    
        There was significant SSA to pure tones in bilateral Te1.0, bilateral Te1.1, and right Te3 ($p < 0.008$, FWE-corrected, see Table \ref{tab:SSAmap}). Significant SSA clusters were present in all anatomical ROIs for FM-sweeps ($p < 0.03$, FWE-corrected, see Table \ref{tab:SSAmap}). SSA magnitude was similar across all anatomical ROIs and both sound modalities.
        
\begin{figure}[h!]
            \centering
            \includegraphics[scale=0.16]{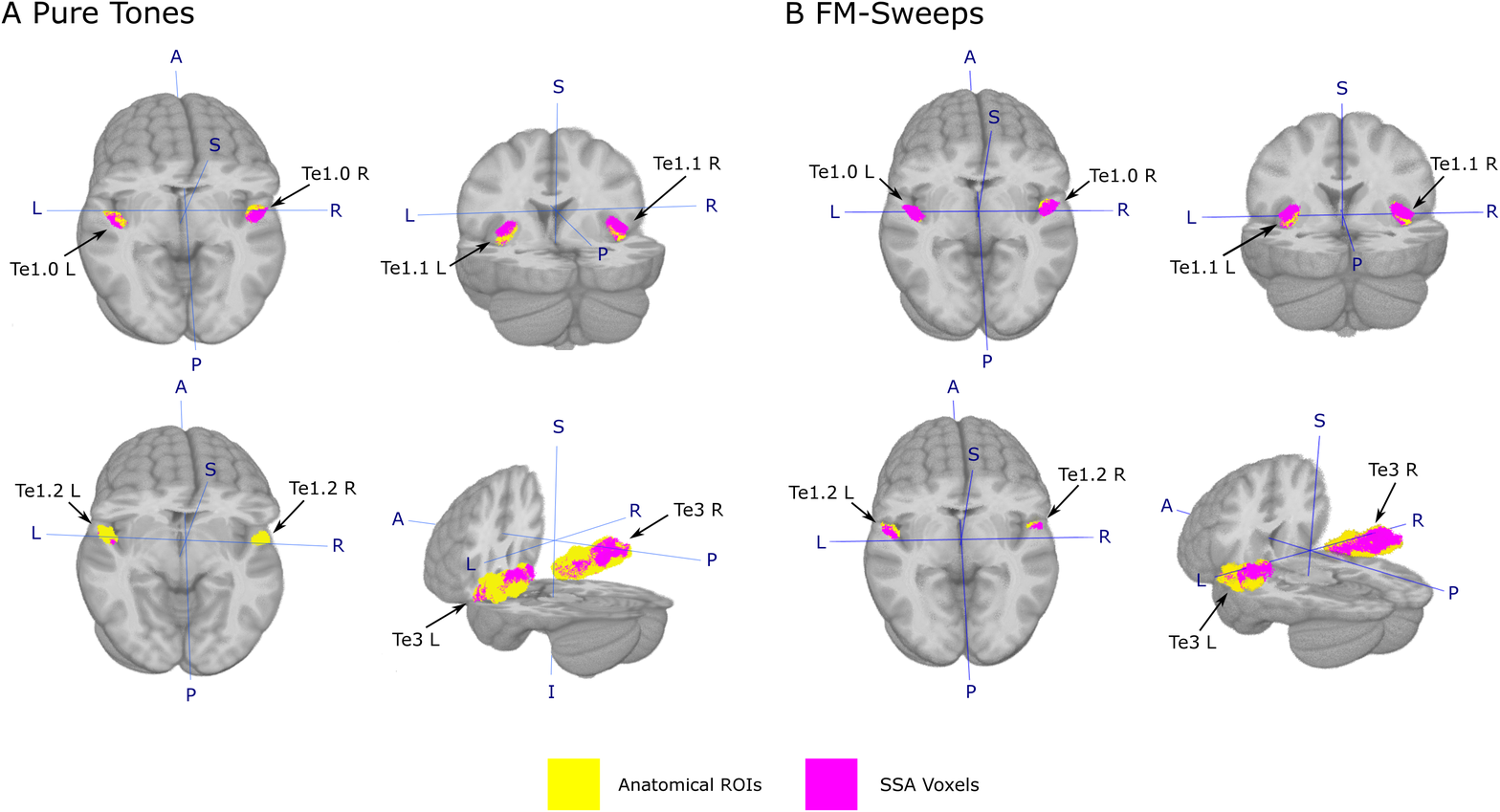}
            \caption[3D View of SSA Voxels]{3D view of voxels showing significant ($p < 0.05$, FDR-corrected) SSA (pink) within all anatomical ROIs (yellow). For both sound modalities, SSA voxels formed spatially coherent fields. A: pure tones; B: FM-sweeps.}
            \label{fig:ssaROIS3D}
        \end{figure}

        Significant SSA clusters formed coherent fields, indicating a systematic spatial encoding pattern. For pure tones, the localization of SSA voxels was lateral within bilateral Te1.0, superior within bilateral Te1.1, and predominantly posterior in right Te3 (Figure \ref{fig:ssaROIS3D}A). For FM-sweeps, the majority of voxels in bilateral Te1.0 and Te1.1 showed significant SSA. SSA voxels in bilateral Te1.2 were localized posterolaterally. In Te3, SSA voxels were mostly found in posterior areas, mirroring the findings from the pure tone experiment (Figure \ref{fig:ssaROIS3D}B).
        
        \begin{table}[h!]
            \centering
            \begin{tabular}{|cc|ccc|ccc|}
                \hline
                 & & & Pure Tones & & & FM-Sweeps &\\
                \hline
                 Contrast  &   ROI   & Size &   Coordinates   & Peak $p$ & Size &   Coordinates   & Peak $p$ \\
                \hline
                Adaptation & Te1.0 L & 539  & $[-48,-25,  9]$ &   0.007  &  950 & $[-42,-20,  5]$ &  0.008   \\
                           & Te1.0 R & 759  & $[ 54,-16,  7]$ &   0.003  & 1085 & $[ 54,-15,  4]$ &  0.009   \\
                           & Te1.1 L & 920  & $[-37,-33, 14]$ &   0.003  & 1145 & $[ 39,-23,  5]$ &  0.008   \\
                           & Te1.1 R & 1374 & $[ 35,-29, 18]$ &   0.003  & 1528 & $[-50,-11, -1]$ &  0.009   \\
                           & Te1.2 L & 118  & $[-48,-13,  2]$ &   0.008  &  715 & $[ 51, -4, -4]$ &  0.02    \\
                           & Te1.2 R & 119s & $[ 57, -3, -6]$ &   0.07   &  730 & $[ 51, -4, -4]$ &  0.01    \\
                           & Te3 L   & 4002 & $[-65,-19,  9]$ &   0.02   & 4747 & $[-58,-14, -4]$ &  0.02    \\
                           & Te3 R   & 3688 & $[ 63,-22,  0]$ &   0.008  & 4200 & $[ 64,-28,  4]$ &  0.02    \\
                \hline  
                  Deviant  & Te1.0 L & 788  & $[-48,-25, 10]$ &   0.003  &  908 & $[-48,-22,  7]$ &  0.008   \\
                 Detection & Te1.0 R & 1042 & $[ 54,-16,  7]$ &   0.002  & 1015 & $[ 51,-18,  5]$ &  0.008   \\
                           & Te1.1 L & 1000 & $[-38,-32, 13]$ &   0.002  & 1114 & $[-38,-24,  8]$ &  0.008   \\
                           & Te1.1 R & 1532 & $[ 38,-22,  6]$ &   0.003  & 1424 & $[ 48,-23,  7]$ &  0.008   \\
                           & Te1.2 L &  619 & $[-53,  2, -3]$ &   0.003  &  631 & $[-51,-11, -1]$ &  0.008   \\
                           & Te1.2 R &  462 & $[ 54, -1, -3]$ &   0.02   &  547 & $[ 57, -4, -4]$ &  0.008   \\
                           & Te3 L   & 3128 & $[-62,-21,  9]$ &   0.01   & 3954 & $[-66,-19,  1]$ &  0.03    \\
                           & Te3 R   & 3239 & $[ 65,-22,  1]$ &   0.007  & 3901 & $[ 63,-28,  4]$ &  0.02    \\
                \hline  
                    SSA    & Te1.0 L &  499 & $[-48,-25,  9]$ &   0.006  &  906 & $[-48,-22,  8]$ &  0.008   \\
                           & Te1.0 R &  748 & $[ 54,-16,  7]$ &   0.002  &  995 & $[ 51,-18,  6]$ &  0.007   \\
                           & Te1.1 L &   88 & $[-37,-33, 14]$ &   0.003  & 1092 & $[-38,-24,  8]$ &  0.007   \\
                           & Te1.1 R & 1372 & $[ 35,-30, 18]$ &   0.005  & 1413 & $[ 41,-28, 14]$ &  0.007   \\
                           & Te1.2 L &   82 & $[-48,-13,  3]$ &   0.06   &  570 & $[-51, -9, -1]$ &  0.02    \\
                           & Te1.2 R &    0 &        -        &     -    &  531 & $[ 51, -4, -4]$ &  0.01    \\
                           & Te3 L   & 2435 & $[-63,-20, 10]$ &   0.05   & 3586 & $[-63,-15, -3]$ &  0.02    \\
                           & Te3 R   & 2439 & $[ 65,-22,  1]$ &   0.007  & 3351 & $[ 66,-20,  6]$ &  0.02    \\
                \hline
            \end{tabular}
            \caption[Peak Adaptation, Deviant Detection, and SSA]{\textbf{Cluster sizes, MNI peak coordinates (mm), and peak-level $p$-values for adaptation, deviant detection, and SSA to pure tones and FM-sweeps.} Voxel-wise $p$-values were FDR-corrected in each ROI and thresholded at $\alpha = 0.05$. Peak-level $p$-values were corrected for the number of voxels per ROI and the total number of comparisons using Bonferroni- and Holm-Bonferroni-correction, respectively. Maps showing voxels in which the respective contrasts were significant are provided in Supplementary Figure \ref{fig:ssaROIS2D} for pure tones and in Supplementary Figure \ref{fig:sweepsssaROIS2D} for FM-sweeps.} 
            \label{tab:SSAmap}
        \end{table}
        
    \subsection*{Magnitude of SSA across cortical fields for pure tones and FM-sweeps}

\begin{figure}[h!]
            \centering
            \includegraphics[scale=0.48]{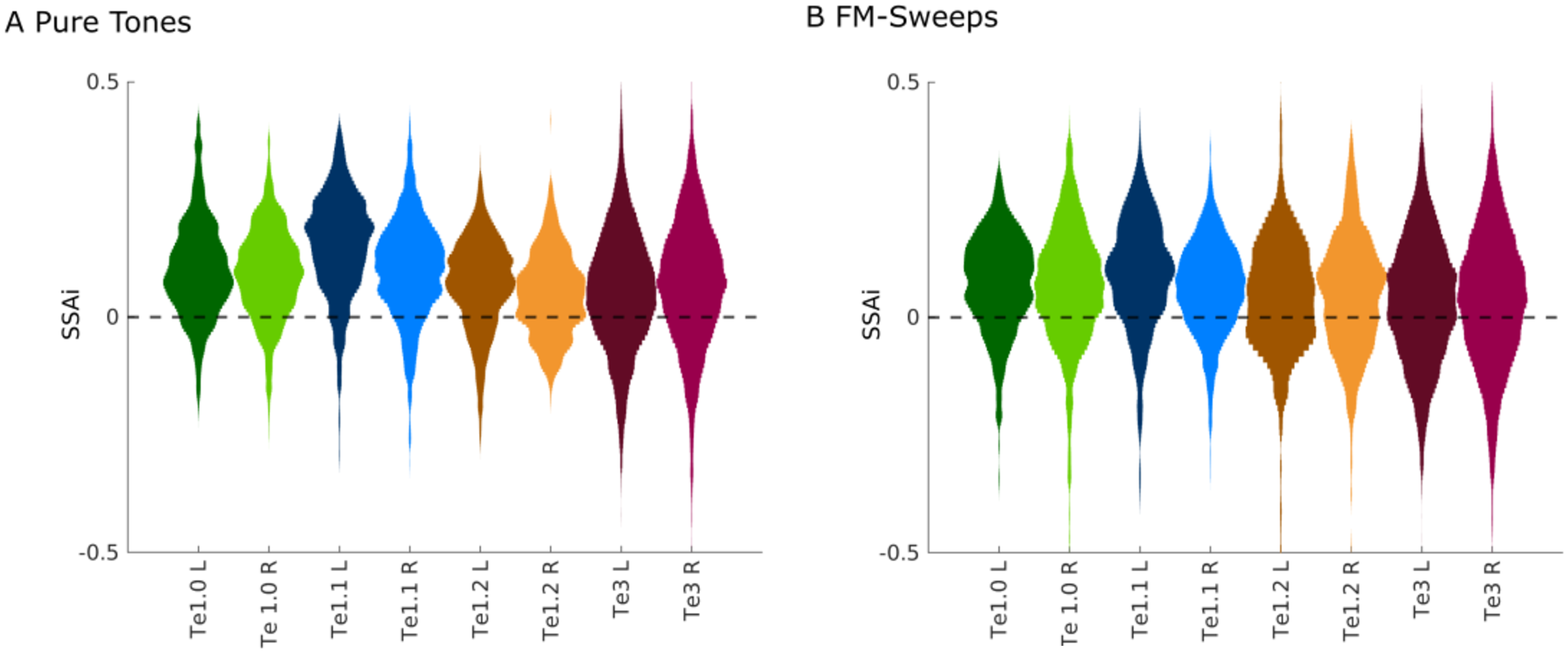}%{FiguresDraft3/SI_dists_anat_comb.png}
            \caption[SSA MAgnitude Distributions]{\textbf{Distributions of SSA magnitude in all eight anatomical ROIs}. We found very similar distributions in all anatomical ROIs and both sound modalities. A: pure tones; B: FM-sweeps.}
            \label{fig:SSAmagnitudeHist}
        \end{figure}

        The SSA magnitude $iSSA$ was similarly distributed across all anatomical ROIs and both sound modalities (Figure \ref{fig:SSAmagnitudeHist}). The topographic distributions of $iSSA$ for pure tones and FM-sweeps showed a slight but significant positive correlation in Te1.0 L, Te1.1 L, and bilateral Te3 ($r \geq 0.04$,$p < 0.01$, corrected for 8 comparisons; Table \ref{tab:SSAmagnitudeCorr}).

        \begin{table}[h!]
            \centering
            \begin{tabular}{ccccccccc}
                     & Te1.0 L & Te1.0 R & Te1.1 L & Te1.1 R & Te1.2 L & Te1.2 R &  Te3 L  & Te3 R \\
                 \hline
                 $r$ &  0.14   &   0.03  &  0.10   &  -0.01  &  -0.21  &  -0.04  &  0.04   &  0.09 \\
                 $p$ & 5.95e-05 &   1  & 4.20e-03 &   1  & 7.00e-12 &   1  & 1.80e-03 & 7.84.0e-17 \\
                \hline
            \end{tabular}
            \caption[SSA Magnitude: Correlations Across Modalities]{\textbf{Correlation between the voxel-wise SSA magnitude for pure tones and FM-sweeps in each anatomical ROI.} $r$: Pearson correlation coefficient. We corrected $p$-values for eight comparisons using Bonferroni-correction.}
            \label{tab:SSAmagnitudeCorr}
        \end{table}
        %\clearpage

    \subsection*{Subjective expectations drive responses to pure tone and FM-sweep deviants}
    
        To test whether subjective expectations modulated the responses to deviants in different positions, we examined the relationship between deviant predictability and deviant responses. Generally, beta estimates qualitatively decreased with increasing deviant predictability in both sound modalities in accordance with the predictive coding hypothesis. We corroborated the effect quantitatively by conducting pair-wise statistical comparisons between the responses to each deviant position at the group-level (Supplementary Tables \ref{tab:groupSignrankPT} and \ref{tab:groupSignrankPT}). In most SSA ROIs, deviants in position four elicited stronger responses than deviants in positions five and six ($p < 0.03$); deviants in position five elicited stronger responses than deviants in position six ($p > 0.04$; all $p$-values of the pure tone data corrected for 30 comparisons and all $p$-values of the FM-sweep data corrected for 48 comparisons). Responses to deviants in position six were not statistically different from responses to standards after the deviants in all SSA ROIs and both sound modalities, with only one exception. Thus, full predictability eliminated deviant detection responses completely.
        
        These pair-wise differences were further confirmed estimating an LMM at the group-level in each SSA ROI. In line with our hypothesis, we found a significant negative effect of deviant predictability on deviant responses in most SSA ROIs of both sound modalities (Table \ref{tab:LMMGroupROIs}).

       \begin{table}[h!]
           \centering
           \setlength{\tabcolsep}{5pt}
           \begin{tabular}{|cc|cccc|cccc|}
               \hline
                     &       &       \multicolumn{4}{c|}{Pure Tones}       &      \multicolumn{4}{c|}{FM-Sweeps}         \\
                \hline
               ROI   & Name  & $\beta$ &   DF   &   $p$   &       CI       & $\beta$ &    DF   &   $p$   &       CI      \\
               \hline
               Te1.0 & Int.  &   0.89  & 113770 & 2.1e-02 & [ 0.33,  1.45] &   0.87  &  437596 & 8.2e-05 & [ 0.49,  1.25]\\
               L     & Pred. &  -1.13  & 113770 & 4.3e-02 & [-1.89, -0.38] &  -1.27  &  437596 & 3.8e-04 & [-1.86, -0.67]\\
               \hline
               Te1.0 & Int.  &   0.10  & 170542 & 4.2e-05 & [ 0.57,  1.39] &   0.80  &  480583 & 3.0e-05 & [ 0.47,  1.13]\\
                R    & Pred. &  -1.245 & 170542 & 3.6e-04 & [-1.82, -0.66] &  -1.11  &  480583 & 1.1e-03 & [-1.67, -0.55]\\
               \hline
               Te1.1 & Int.  &   1.22  & 202234 & 8.4e-05 & [ 0.69,  1.04] &   1.04  &  527434 & 9.2e-11 & [ 0.74,  1.34]\\
                L    & Pred. & - 1.52  & 202234 & 4.9e-04 & [-2.24, -0.80] &  -1.47  &  527434 & 8.5e-11 & [-1.89, -1.95]\\
               \hline
               Te1.1 & Int.  &   1.17  & 312814 & 1.4e-05 & [ 0.70,  1.64] &   0.90  &  682477 & 1.3e-11 & [ 0.65,  1.15]\\
               R     & Pred. &  -1.41  & 312814 & 4.6e-04 & [-2.08, -0.74] &  -1.36  &  682477 & 4.2e-10 & [-1.77, -0.96]\\
               \hline
               Te1.2 & Int.  &     -   &    -   &    -    &         -      &   0.69  &  275308 & 3.9e-02 & [ 0.23,  1.14]\\
               L     & Pred. &     -   &    -   &    -    &         -      &  -0.96  &  275308 & 8.2e-02 & [-1.65, -0.27]\\
               \hline
               Te1.2 & Int.  &     -   &    -   &    -    &         -      &   0.49  &  256471 & 6.7e-02 & [ 0.14,  0.84]\\
                R    & Pred. &     -   &    -   &    -    &         -      &  -0.60  &  256471 & 3.9e-01 & [-1.14, -0.06]\\
               \hline
               Te3   & Int.  &     -   &    -   &    -    &         -      &   0.61  & 1732036 & 1.4e-05 & [ 0.36,  0.86]\\
                L    & Pred. &     -   &    -   &    -    &         -      &  -0.75  & 1732036 & 4.4e-04 & [-1.11, -0.40]\\
               \hline
               Te3   & Int.  &   0.70  & 556090 & 3.8e-02 & [ 0.23,  1.16] &   0.70  & 1618531 & 3.1e-05 & [ 0.41,  1.00]\\
               R     & Pred. &   -0.8  & 556090 & 1.4e-01 & [-1.46, -0.19] &  -0.83  & 1618531 & 2.6e-03 & [-1.26, -0.39]\\
               \hline
           \end{tabular}
           \caption[Group-Level LMM]{\textbf{Fixed effect coefficients of the group-level LMM}. Pure tone model: beta $\sim$~1 + predictability + (1 + predictability\textbar run) + (1 + predictability\textbar participant); FM-sweep model: beta $\sim$~1 + predictability + (1 + predictability\textbar session) + (1 + predictability\textbar run) + (1 + predictability\textbar participant). \emph{Int.}: Intercept; \emph{Pred.}: fixed effect regressor of the model, predictability of the deviants; \emph{DF}: degrees of freedom; $p$: $p$-value of the $t$-test testing for the equality of the coefficient to zero; \emph{CI}: limits of the confidence interval for the coefficients. All $p$-values were Bonferroni-corrected for $N = 13$ SSA ROIs.}
           \label{tab:LMMGroupROIs}
           \end{table}
       \setlength{\tabcolsep}{6pt}
       
       We could replicate the group-level results in most individual participants for both sound modalities: the correlation between deviant predictability and deviant responses significantly negative in all but one participant for the pure tone data (for those participants $\rho \in [-0.66, -.011]$, all $p < 10^{-7}$, see Supplementary Figure \ref{fig:correlationssubPT}. For FM-sweeps, this was also the case in 16 out of 18 participants (for these participants $\rho \in [-0.68, -0.11]$, all $p < 10^{-91}$, see Supplementary Figure \ref{fig:correlationssubSweeps}). 

    \subsection*{Neural responses to pure tones and FM-sweeps are best explained by predictive coding}
        
\begin{figure}[h!]
            \centering
            \includegraphics[scale=0.19]{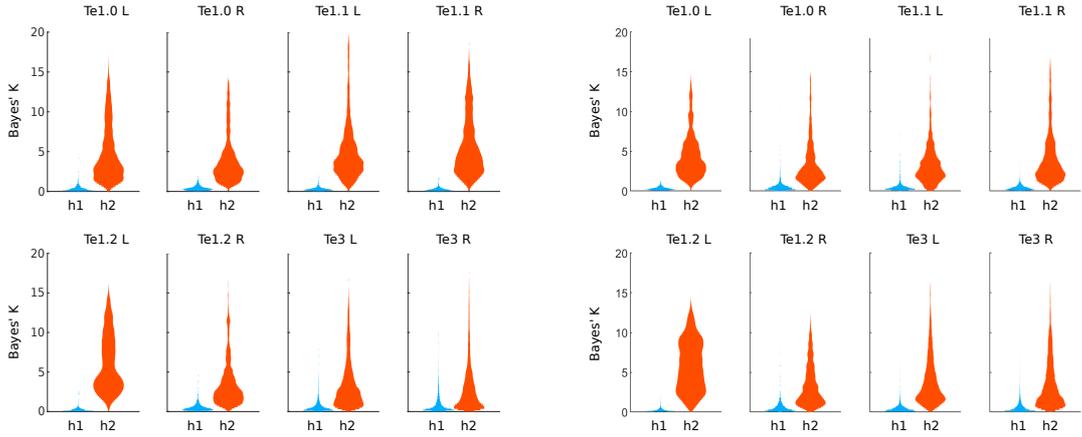}
            \caption[Bayesian Model Comparison: Habituation (h1) vs. prediction error (h2)]{\textbf{Distributions of Bayes' $K$ for the habituation (h1) and prediction error (h2) model in all anatomical ROIs.} The results show that most voxels across ROIs and sound modalities yielded evidence in favor of h2. A: pure tone data set; B: FM-sweep data set.}
            \label{fig:BMCHabPred}
        \end{figure}

        For both stimulus types, the prediction error model (h2) outperformed the habituation model (h1) in line with our hypothesis (see Figure \ref{fig:BMCHabPred} for the distribution of Bayes' $K$ of both models for each anatomical ROI and stimulus type). Voxel-wise maps of Bayes $K$ for h2 are shown in Supplementary Figure \ref{fig:H2KPT} for pure tones and Supplementary Figure \ref{fig:H2KSweeps} for FM-sweeps. Descriptively, voxels with higher Bayes factor formed spatially coherent fields, indicating a functional topographic organization.
            
        For the pure tone data, we found evidence in favor of h2 (posterior density $P > 0.5$) in 96\% of voxels in Te1.0 L, 94\% in Te1.0 R, 97\% in Te1.1 L, 99\% in Te1.1 R, 99\% in Te1.2 L, 90\% in Te1.2 R, 82\% in Te3 L, and 72\% in Te3 R. For the FM-sweeps, we also found evidence in favor of the predictability model in a majority of voxels of all anatomical ROIs (98\% of voxels in Te1.0 L, 89\% in Te1.0 R, 90\% in Te1.1 L, 94\% in Te1.1 R, 97\% in Te1.2 L, 83\% in Te1.2 R, 91\% in Te3 L, and 82\% in Te3 R).

        %\clearpage
        
    \subsection*{Similar neural populations encode prediction error to pure tones and FM-sweeps in some but not all cortical fields}
    
        Computing correlations between the $K$-maps of both stimulus types for both models, we found significant and positive correlation coefficients for h1 in all ROIs but left Te1.0 ($r > 0.15$, $p < {10}^-4$, Bonferroni-corrected for 8 comparisons) and for h2 in left Te1.1, left Te1.2 and bilateral Te3. ($r > 0.14$, $p < {10}^-5$, Bonferroni-corrected for 8 comparisons; Table~\ref{tab:h1h2Corr}).
    
        \begin{table}[h!]
            \centering
            \begin{tabular}{ccccccccc}
            h1\\
            \hline
            & Te1.0 L & Te1.0 R & Te1.1 L & Te1.1 R & Te1.2 L & Te1.2 R & Te3 L & Te3 R \\ [0.5ex]
             \hline
             $r$ & -0.0035 & 0.1571 & 0.3653 & 0.1608 & 0.2462 & 0.1598 & 0.2366 & 0.2921\\
             $p$ & 1 & 3.21e-07 & 3.66e-41 & 5.86e-10 & 6.16e-06 & 2.85e-05 & 4.13e-90 & 1.173e-162\\
            \hline
            \\
            h2\\
            \hline
            & Te1.0 L & Te1.0 R & Te1.1 L & Te1.1 R & Te1.2 L & Te1.2 R & Te3 L & Te3 R \\ [0.5ex]
            \hline
            $r$ & -0.0042 & -0.0997 & 0.3675 & 0.0459 & 0.14556 & 0.0807 & 0.2400 & 0.2978\\
            $p$ & 1 & 4.2e-03 & 1.07e-41 & 0.52 & 8.07e-06 & 0.16 & 9.97e-93 & 2.11e-169\\
            \hline
            \end{tabular}
            \caption[]{\textbf{Correlations between the Bayes factor maps of both stimulus types.} $r$: Pearson correlation coefficient. We Bonferroni-corrected all $p$-values associated with each hypothesis for eight comparisons.}
            \label{tab:h1h2Corr}
        \end{table}

\section*{Discussion}

    %\subsection*{Summary}
    
        The main tenet of the predictive coding framework is that sensory neurons encode prediction error with respect to an internal generative model of the world \cite{rao1999predictive, mumford1992computational, friston2003learning, spratling2017review}. Here, we have tested two hypotheses: first, whether prediction error in different fields of the human AC is encoded with respect to a generative model informed by the task instructions and the encoding of subjective expectations. Second, whether the encoding of prediction error is equivalent for two stimulus types: pure tones and FM-sweeps. We conducted two fMRI experiments, one using each stimulus types. We used a modified oddball paradigm where we manipulated participants' expectations independently of local stimulus statistics. 
        There were three key findings: First, we found significant SSA to pure tones and FM-sweeps across cortical fields, indicating that neural populations in the AC adapt to both, pure tones and FM-sweeps. Second, we found that neural adaptation was driven by the participant's expectations, demonstrating that SSA reflects prediction error with respect to a generative model that is informed by the task instructions.  Third, we found that the populations encoding prediction error to pure tones and FM-sweeps overlap significantly in bilateral Te3, left Te1.1, and left Te1.2, demonstrating that, at least in those fields, both stimulus types share a common mechanism for the computation of prediction error. Together, our results suggest that predictive coding is the general encoding mechanism of acoustic information in AC.
        
    \paragraph{}%\subsection*{Neural populations of AC adapt to FM-sweeps}
        
        Our results are the first robust evidence for prediction error encoding of FM-sweeps in human AC. Previous studies investigating prediction error encoding of FM-sweeps reported mixed results. Three studies reported a significant MMN to deviating FM-sweeps \cite{cornella2013regularity, kung2020minimum, hsieh2021interaction}, but did not localize the source of the response. One study investigated sources in the AC, but found no significant result \cite{altmann2011repetition}. Three other studies reported increasing neuromagnetic responses to repeated FM-sweeps \cite{heinemann2010repetition, heinemann2011auditory, okamoto2017modulation}, in direct contradiction with predictive coding. One of these studies \cite{okamoto2017modulation} reported the effect specifically in the AC. Differences in the stimulus features and ISIs used in these studies might have contributed to these contradictory results \cite{heinemann2011auditory}.
        
        In animal research, SSA to FM direction was reported in A1 of rats \cite{klein2014stimulus}. SSA to FM is also present in the inferior colliculus of bats \cite{thomas2012stimulus}. However, since FM sounds are essential for echolocation in bats \cite{yue2007response}, it is unclear whether this result is transferable to humans. 
        
    \paragraph{}%\subsection*{Common encoding mechanisms for pure tones and FM-sweeps.}
        
        Our results indicate that similar neural populations in the human AC encode prediction error to pure tones and FM-sweeps: First, all SSA clusters formed spatially coherent fields across cortical ROIs for both pure tones and FM-sweeps, suggesting a systematic functional organization of SSA in the AC. Second, we showed that similar neural populations encode prediction error to pure tones and FM-sweeps in left Te1.1, left Te1.2, and bilateral Te3. Given that the data was collected using different scanners and different participants, our results suggest that prediction error is a stimulus-independent encoding mechanism in the AC.
        
        Our results further showed that overlapping neural populations encode prediction error to two out of three auditory information-bearing units (IBUs): pure tones and FM-sweeps. The IBUs are the basic auditory information-bearing elements that conform information-carrying acoustic signals \cite{Suga2012}. The result that overlapping neural populations encode prediction error to two out of three IBUs may suggest that the same predictive encoding mechanisms might underlie the processing of all information-carrying acoustic signals in the AC.

    \paragraph{}% \paragraph{Systematic differences between fields}
        
        Animal studies have reported stronger SSA in secondary compared to primary auditory cortical areas \cite{nieto2016topographic, parras2017neurons}. However, recent evidence suggests that this pattern might be more complex than previously assumed: using single-cell recordings across auditory fields in rats, prediction error was largest in the posterior auditory field---a secondary auditory area. Primary areas and the secondary suprarhinal auditory field, on the other hand, showed stronger effects of stimulus repetition \cite{parras2021posterior}. 
        
        If the functional organization of the AC was similar in rodents and humans, we would expect stronger SSA in Te1.2 and Te3 compared to the remaining ROIs. However, we found similar SSA magnitudes across fields. These discrepancies may stem from a poor correspondence between auditory fields in humans and rodents \cite{morosan2001human, moerel2014anatomical}. However, our previous results showed that the distributions of SSA in the subcortical auditory pathway, closely replicated across mammals \cite{glendenning1998comparative}, also differed between humans and rodents \cite{tabas2020abstract, tabas2021fast}. 
        
        Our results did however suggest a potential differential role of primary and higher-level human auditory cortical areas: First, we did not find significant SSA clusters in bilateral Te1.2 for pure tones; second, the proportion of SSA-voxels was comparably lower in Te3 for both stimulus types. Last, we found evidence for a reliable prediction error encoding topography across stimulus types in bilateral Te3, but not in Te1.0, and only unilaterally in Te1.1 and Te1.2. Together, these results suggest that the encoding of prediction error is stronger and potentially more modality-specific in primary auditory cortical areas, and more general in higher-order regions.
        
    \paragraph{}% \subsection*{The role of high-level expectations in computing prediction errors along the auditory hierarchy}

        Formulations of the predictive coding framework disagree on whether predictions from generative model units inform prediction errors only at the immediate lower stage \cite{spratling2017review, keller2018predictive} or also at subsequent stages of the processing hierarchy (see \cite{tabas2021adjudicating} for a review of the empirical evidence on both standpoints). MMN studies showed that prediction error is elicited with respect to high-level expectations; namely: by the violation of complex statistical regularities (see \cite{paavilainen2013mismatch} for review), the omission of expected sounds \cite{bendixen2009heard, wacongne2011evidence, chennu2016silent}, and abstract expectations about the occurrence of deviating sounds \cite{wacongne2011evidence}. However, since the generators of the MMN are partly located in the frontal cortex \cite{paavilainen2013mismatch}, MMN research is not suitable to clarify whether subjective expectations are used to compute prediction errors at lower levels of the auditory processing hierarchy.
        
        We found prediction error encoding to be the dominant encoding principle for both stimulus types in all anatomical ROIs. This result suggests that high-level predictions informed by the task instructions, putatively computed in regions at higher processing stages than the sensory cortices, are used to compute prediction errors in the primary AC. We had previously shown that these same predictions are also used to compute prediction error in the human auditory midbrain and thalamus \cite{tabas2020abstract, tabas2021fast}. Previous studies also showed that prediction error in the AC was computed with respect to language-specific expectations (e.g., \cite{blank2016prediction, ylinen2016predictive, blank2018neural, heilbron2021hierarchy}). Together, the empirical evidence supports the hypothesis that high-level predictions are used to compute prediction errors along the entire processing hierarchy. 
        
    \paragraph{}%\paragraph{Does SSA in AC reflect bottom-up input from subcortical areas?}
        
        We have previously shown that SSA and prediction error encoding of pure tones and FM-sweeps are already present in the auditory thalamus and midbrain \cite{tabas2020abstract, tabas2021fast}. Since auditory cortical areas receive direct bottom-up input from the auditory thalamus \cite{malmierca2015auditory}, our results here might simply reflect ascending input from prediction error units in subcortical structures. On the other hand, subcortical SSA and prediction error signals might as well reflect top-down cortical input. Animal studies have shown that SSA in the auditory midbrain and thalamus persists under deactivation of the AC \cite{antunes2011effect, anderson2013effect}, at least in regions with strong SSA  \cite{bauerle2011stimulus}. Further work is needed to clarify the interplay of bottom-up and top-down signalling in the computation of prediction error.
        
    \paragraph{}%\subsection*{Conclusion and Significance}
        
        Our results suggest that predictive coding is the general mechanism underlying the encoding of information-bearing acoustic signals in AC. Impaired predictive processes in AC have been linked to speech processing disorders and clinical conditions such as developmental dyslexia (e.g., \cite{plakas2013impaired, neuhoff2012evidence, gu2020auditory}), stuttering \cite{daliri2015electrophysiological}, autism spectrum disorder \cite{van2017predictive}, psychosis \cite{fryer2020deficits}, or schizophrenia \cite{perez2014automatic}. Investigating how predictive coding is implemented in the human AC is essential for a mechanistic understanding of these disorders.

\cleardoublepage

\bibliography{bibliography}

\begin{thebibliography}{10}

\bibitem{de2018expectations}
F.~P. De~Lange, M.~Heilbron, and P.~Kok, ``How do expectations shape
  perception?,'' {\em Trends in cognitive sciences}, vol.~22, no.~9,
  pp.~764--779, 2018.

\bibitem{leonard2016perceptual}
M.~K. Leonard, M.~O. Baud, M.~J. Sjerps, and E.~F. Chang, ``Perceptual
  restoration of masked speech in human cortex,'' {\em Nature communications},
  vol.~7, no.~1, pp.~1--9, 2016.

\bibitem{chalk2010rapidly}
M.~Chalk, A.~R. Seitz, and P.~Seri{\`e}s, ``Rapidly learned stimulus
  expectations alter perception of motion,'' {\em Journal of vision}, vol.~10,
  no.~8, pp.~2--2, 2010.

\bibitem{chambers2017prior}
C.~Chambers, S.~Akram, V.~Adam, C.~Pelofi, M.~Sahani, S.~Shamma, and
  D.~Pressnitzer, ``Prior context in audition informs binding and shapes simple
  features,'' {\em Nature communications}, vol.~8, no.~1, pp.~1--11, 2017.

\bibitem{sterzer2008believing}
P.~Sterzer, C.~Frith, and P.~Petrovic, ``Believing is seeing: expectations
  alter visual awareness,'' {\em Current Biology}, vol.~18, no.~16,
  pp.~R697--R698, 2008.

\bibitem{lange2009brain}
K.~Lange, ``Brain correlates of early auditory processing are attenuated by
  expectations for time and pitch,'' {\em Brain and cognition}, vol.~69, no.~1,
  pp.~127--137, 2009.

\bibitem{rao1999predictive}
R.~P. Rao and D.~H. Ballard, ``Predictive coding in the visual cortex: a
  functional interpretation of some extra-classical receptive-field effects,''
  {\em Nature neuroscience}, vol.~2, no.~1, pp.~79--87, 1999.

\bibitem{mumford1992computational}
D.~Mumford, ``On the computational architecture of the neocortex,'' {\em
  Biological cybernetics}, vol.~66, no.~3, pp.~241--251, 1992.

\bibitem{friston2003learning}
K.~Friston, ``Learning and inference in the brain,'' {\em Neural Networks},
  vol.~16, no.~9, pp.~1325--1352, 2003.

\bibitem{spratling2017review}
M.~W. Spratling, ``A review of predictive coding algorithms,'' {\em Brain and
  cognition}, vol.~112, pp.~92--97, 2017.

\bibitem{nieto2016topographic}
J.~Nieto-Diego and M.~S. Malmierca, ``Topographic distribution of
  stimulus-specific adaptation across auditory cortical fields in the
  anesthetized rat,'' {\em PLoS biology}, vol.~14, no.~3, p.~e1002397, 2016.

\bibitem{rubin2016representation}
J.~Rubin, N.~Ulanovsky, I.~Nelken, and N.~Tishby, ``The representation of
  prediction error in auditory cortex,'' {\em PLoS computational biology},
  vol.~12, no.~8, p.~e1005058, 2016.

\bibitem{parras2017neurons}
G.~G. Parras, J.~Nieto-Diego, G.~V. Carbajal, C.~Vald{\'e}s-Baizabal,
  C.~Escera, and M.~S. Malmierca, ``Neurons along the auditory pathway exhibit
  a hierarchical organization of prediction error,'' {\em Nature
  communications}, vol.~8, no.~1, pp.~1--17, 2017.

\bibitem{perez2021deviance}
D.~P{\'e}rez-Gonz{\'a}lez, G.~G. Parras, C.~J. Morado-D{\'\i}az,
  C.~Aedo-S{\'a}nchez, G.~V. Carbajal, and M.~S. Malmierca, ``Deviance
  detection in physiologically identified cell types in the rat auditory
  cortex,'' {\em Hearing Research}, vol.~399, p.~107997, 2021.

\bibitem{cacciaglia2019auditory}
R.~Cacciaglia, J.~Costa-Faidella, K.~Zarnowiec, S.~Grimm, and C.~Escera,
  ``Auditory predictions shape the neural responses to stimulus repetition and
  sensory change,'' {\em NeuroImage}, vol.~186, pp.~200--210, 2019.

\bibitem{zvyagintsev2020auditory}
M.~Zvyagintsev, J.~Zweerings, P.~Sarkheil, S.~Bergert, H.~Baqapuri, I.~Neuner,
  A.~J. Gaebler, and K.~Mathiak, ``Auditory mismatch processing: role of
  paradigm and stimulus characteristics as detected by fmri,'' {\em Biological
  Psychology}, p.~107887, 2020.

\bibitem{ulanovsky2003processing}
N.~Ulanovsky, L.~Las, and I.~Nelken, ``Processing of low-probability sounds by
  cortical neurons,'' {\em Nature neuroscience}, vol.~6, no.~4, pp.~391--398,
  2003.

\bibitem{eytan2003selective}
D.~Eytan, N.~Brenner, and S.~Marom, ``Selective adaptation in networks of
  cortical neurons,'' {\em Journal of Neuroscience}, vol.~23, no.~28,
  pp.~9349--9356, 2003.

\bibitem{mill2011neurocomputational}
R.~Mill, M.~Coath, T.~Wennekers, and S.~L. Denham, ``A neurocomputational model
  of stimulus-specific adaptation to oddball and markov sequences,'' {\em PLoS
  Comput Biol}, vol.~7, no.~8, p.~e1002117, 2011.

\bibitem{wang2014stimulus}
H.~Wang, Y.-F. Han, Y.-S. Chan, and J.~He, ``Stimulus-specific adaptation at
  the synapse level in vitro,'' {\em PloS one}, vol.~9, no.~12, p.~e114537,
  2014.

\bibitem{tabas2021adjudicating}
A.~Tabas {\em et~al.}, ``Adjudicating between local and global architectures of
  predictive processing in the subcortical auditory pathway,'' {\em Frontiers
  in Neural Circuits}, vol.~15, p.~12, 2021.

\bibitem{carbajal2018neuronal}
G.~V. Carbajal and M.~S. Malmierca, ``The neuronal basis of predictive coding
  along the auditory pathway: from the subcortical roots to cortical deviance
  detection,'' {\em Trends in hearing}, vol.~22, p.~2331216518784822, 2018.

\bibitem{tabas2020abstract}
A.~Tabas, G.~Mihai, S.~Kiebel, R.~Trampel, and K.~von Kriegstein, ``Abstract
  rules drive adaptation in the subcortical sensory pathway,'' {\em Elife},
  vol.~9, p.~e64501, 2020.

\bibitem{costa2011interactions}
J.~Costa-Faidella, T.~Baldeweg, S.~Grimm, and C.~Escera, ``Interactions between
  “what” and “when” in the auditory system: temporal predictability
  enhances repetition suppression,'' {\em Journal of Neuroscience}, vol.~31,
  no.~50, pp.~18590--18597, 2011.

\bibitem{todorovic2011prior}
A.~Todorovic, F.~van Ede, E.~Maris, and F.~P. de~Lange, ``Prior expectation
  mediates neural adaptation to repeated sounds in the auditory cortex: an meg
  study,'' {\em Journal of Neuroscience}, vol.~31, no.~25, pp.~9118--9123,
  2011.

\bibitem{cornella2012detection}
M.~Cornella, S.~Leung, S.~Grimm, and C.~Escera, ``Detection of simple and
  pattern regularity violations occurs at different levels of the auditory
  hierarchy,'' {\em PLoS One}, vol.~7, no.~8, p.~e43604, 2012.

\bibitem{todorovic2012repetition}
A.~Todorovic and F.~P. de~Lange, ``Repetition suppression and expectation
  suppression are dissociable in time in early auditory evoked fields,'' {\em
  Journal of Neuroscience}, vol.~32, no.~39, pp.~13389--13395, 2012.

\bibitem{lecaignard2015implicit}
F.~Lecaignard, O.~Bertrand, G.~Gimenez, J.~Mattout, and A.~Caclin, ``Implicit
  learning of predictable sound sequences modulates human brain responses at
  different levels of the auditory hierarchy,'' {\em Frontiers in Human
  Neuroscience}, vol.~9, p.~505, 2015.

\bibitem{durschmid2016hierarchy}
S.~D{\"u}rschmid, E.~Edwards, C.~Reichert, C.~Dewar, H.~Hinrichs, H.-J. Heinze,
  H.~E. Kirsch, S.~S. Dalal, L.~Y. Deouell, and R.~T. Knight, ``Hierarchy of
  prediction errors for auditory events in human temporal and frontal cortex,''
  {\em Proceedings of the National Academy of Sciences}, vol.~113, no.~24,
  pp.~6755--6760, 2016.

\bibitem{phillips2016convergent}
H.~N. Phillips, A.~Blenkmann, L.~E. Hughes, S.~Kochen, T.~A. Bekinschtein,
  J.~B. Rowe, {\em et~al.}, ``Convergent evidence for hierarchical prediction
  networks from human electrocorticography and magnetoencephalography,'' {\em
  cortex}, vol.~82, pp.~192--205, 2016.

\bibitem{shalgi2007direct}
S.~Shalgi and L.~Y. Deouell, ``Direct evidence for differential roles of
  temporal and frontal components of auditory change detection,'' {\em
  Neuropsychologia}, vol.~45, no.~8, pp.~1878--1888, 2007.

\bibitem{deouell2007frontal}
L.~Y. Deouell, ``The frontal generator of the mismatch negativity revisited,''
  {\em Journal of Psychophysiology}, vol.~21, no.~3-4, pp.~188--203, 2007.

\bibitem{berlot2018mapping}
E.~Berlot, E.~Formisano, and F.~De~Martino, ``Mapping frequency-specific tone
  predictions in the human auditory cortex at high spatial resolution,'' {\em
  Journal of Neuroscience}, vol.~38, no.~21, pp.~4934--4942, 2018.

\bibitem{lee2011classification}
C.~C. Lee and S.~M. Sherman, ``On the classification of pathways in the
  auditory midbrain, thalamus, and cortex,'' {\em Hearing research}, vol.~276,
  no.~1-2, pp.~79--87, 2011.

\bibitem{hu2003functional}
B.~Hu, ``Functional organization of lemniscal and nonlemniscal auditory
  thalamus,'' {\em Experimental Brain Research}, vol.~153, no.~4, pp.~543--549,
  2003.

\bibitem{antunes2010stimulus}
F.~M. Antunes, I.~Nelken, E.~Covey, and M.~S. Malmierca, ``Stimulus-specific
  adaptation in the auditory thalamus of the anesthetized rat,'' {\em PLoS
  One}, vol.~5, no.~11, p.~e14071, 2010.

\bibitem{antunes2011effect}
F.~M. Antunes and M.~S. Malmierca, ``Effect of auditory cortex deactivation on
  stimulus-specific adaptation in the medial geniculate body,'' {\em Journal of
  Neuroscience}, vol.~31, no.~47, pp.~17306--17316, 2011.

\bibitem{duque2012topographic}
D.~Duque, D.~P{\'e}rez-Gonz{\'a}lez, Y.~A. Ayala, A.~R. Palmer, and M.~S.
  Malmierca, ``Topographic distribution, frequency, and intensity dependence of
  stimulus-specific adaptation in the inferior colliculus of the rat,'' {\em
  Journal of Neuroscience}, vol.~32, no.~49, pp.~17762--17774, 2012.

\bibitem{duque2014modulation}
D.~Duque, M.~S. Malmierca, and D.~M. Caspary, ``Modulation of stimulus-specific
  adaptation by gabaa receptor activation or blockade in the medial geniculate
  body of the anaesthetized rat,'' {\em The Journal of physiology}, vol.~592,
  no.~4, pp.~729--743, 2014.

\bibitem{ayala2015differences}
Y.~A. Ayala, A.~Udeh, K.~Dutta, D.~Bishop, M.~S. Malmierca, and D.~L. Oliver,
  ``Differences in the strength of cortical and brainstem inputs to ssa and
  non-ssa neurons in the inferior colliculus,'' {\em Scientific reports},
  vol.~5, no.~1, pp.~1--17, 2015.

\bibitem{Suga2012}
N.~Suga, ``Basic acoustic patterns and neural mechanisms shared by humans and
  animals for auditory perception,'' in {\em Listening to Speech}, Psychology
  Press, 2012.

\bibitem{divenyi2009perception}
P.~Divenyi, ``Perception of complete and incomplete formant transitions in
  vowels,'' {\em The Journal of the Acoustical Society of America}, vol.~126,
  no.~3, pp.~1427--1439, 2009.

\bibitem{liberman1967perception}
A.~M. Liberman, F.~S. Cooper, D.~P. Shankweiler, and M.~Studdert-Kennedy,
  ``Perception of the speech code.,'' {\em Psychological review}, vol.~74,
  no.~6, p.~431, 1967.

\bibitem{malmierca2010structural}
M.~S. Malmierca and T.~A. Hackett, ``Structural organization of the ascending
  auditory pathway,'' {\em The Auditory Brain}, vol.~2, pp.~9--41, 2010.

\bibitem{geis2013intracellular}
R.~Geis and G.~Borst, ``Intracellular responses to frequency modulated tones in
  the dorsal cortex of the mouse inferior colliculus,'' {\em Frontiers in
  neural circuits}, vol.~7, p.~7, 2013.

\bibitem{lui2003frequency}
B.~Lui and J.~Mendelson, ``Frequency modulated sweep responses in the medial
  geniculate nucleus,'' {\em Experimental brain research}, vol.~153, no.~4,
  pp.~550--553, 2003.

\bibitem{issa2017multiscale}
J.~B. Issa, B.~D. Haeffele, E.~D. Young, and D.~T. Yue, ``Multiscale mapping of
  frequency sweep rate in mouse auditory cortex,'' {\em Hearing research},
  vol.~344, pp.~207--222, 2017.

\bibitem{hall2000modulation}
D.~A. Hall, M.~P. Haggard, M.~A. Akeroyd, A.~Q. Summerfield, A.~R. Palmer,
  M.~R. Elliott, and R.~W. Bowtell, ``Modulation and task effects in auditory
  processing measured using fmri,'' {\em Human brain mapping}, vol.~10, no.~3,
  pp.~107--119, 2000.

\bibitem{hart2003amplitude}
H.~C. Hart, A.~R. Palmer, and D.~A. Hall, ``Amplitude and frequency-modulated
  stimuli activate common regions of human auditory cortex,'' {\em Cerebral
  Cortex}, vol.~13, no.~7, pp.~773--781, 2003.

\bibitem{paltoglou2011mapping}
A.~E. Paltoglou, C.~J. Sumner, and D.~A. Hall, ``Mapping feature-sensitivity
  and attentional modulation in human auditory cortex with functional magnetic
  resonance imaging,'' {\em European Journal of Neuroscience}, vol.~33, no.~9,
  pp.~1733--1741, 2011.

\bibitem{altmann2014representation}
C.~F. Altmann and B.~H. Gaese, ``Representation of frequency-modulated sounds
  in the human brain,'' {\em Hearing research}, vol.~307, pp.~74--85, 2014.

\bibitem{heinemann2010repetition}
L.~V. Heinemann, B.~Rahm, J.~Kaiser, B.~H. Gaese, and C.~F. Altmann,
  ``Repetition enhancement for frequency-modulated but not unmodulated sounds:
  a human meg study,'' {\em PLoS One}, vol.~5, no.~12, p.~e15548, 2010.

\bibitem{altmann2011repetition}
C.~F. Altmann, C.~Klein, L.~V. Heinemann, M.~Wibral, B.~H. Gaese, and
  J.~Kaiser, ``Repetition of complex frequency-modulated sweeps enhances
  neuromagnetic responses in the human auditory cortex,'' {\em Hearing
  research}, vol.~282, no.~1-2, pp.~216--224, 2011.

\bibitem{heinemann2011auditory}
L.~V. Heinemann, J.~Kaiser, and C.~F. Altmann, ``Auditory repetition
  enhancement at short interstimulus intervals for frequency-modulated tones,''
  {\em Brain research}, vol.~1411, pp.~65--75, 2011.

\bibitem{okamoto2017modulation}
H.~Okamoto and R.~Kakigi, ``Modulation of auditory evoked magnetic fields
  elicited by successive frequency-modulated (fm) sweeps,'' {\em Frontiers in
  Human Neuroscience}, vol.~11, p.~36, 2017.

\bibitem{cornella2013regularity}
M.~Cornella, S.~Leung, S.~Grimm, and C.~Escera, ``Regularity encoding and
  deviance detection of frequency modulated sweeps: Human middle-and
  long-latency auditory evoked potentials,'' {\em Psychophysiology}, vol.~50,
  no.~12, pp.~1275--1281, 2013.

\bibitem{kung2020minimum}
S.-J. Kung, D.~H. Wu, C.-H. Hsu, and I.-H. Hsieh, ``A minimum temporal window
  for direction detection of frequency-modulated sweeps: A
  magnetoencephalography study,'' {\em Frontiers in Psychology}, vol.~11,
  p.~389, 2020.

\bibitem{hsieh2021interaction}
I.-H. Hsieh and W.-T. Yeh, ``The interaction between timescale and pitch
  contour at pre-attentive processing of frequency-modulated sweeps,'' {\em
  Frontiers in Psychology}, vol.~12, p.~697, 2021.

\bibitem{tabas2021fast}
A.~Tabas, S.~Kiebel, M.~Marxen, and K.~von Kriegstein, ``Fast frequency
  modulation is encoded according to the listener expectations in the human
  subcortical auditory pathway,'' {\em arXiv preprint arXiv:2108.02066}, 2021.

\bibitem{friston1999stochastic}
K.~J. Friston, E.~Zarahn, O.~Josephs, R.~N. Henson, and A.~M. Dale,
  ``Stochastic designs in event-related fmri,'' {\em Neuroimage}, vol.~10,
  no.~5, pp.~607--619, 1999.

\bibitem{gorgolewski2011nipype}
K.~Gorgolewski, C.~D. Burns, C.~Madison, D.~Clark, Y.~O. Halchenko, M.~L.
  Waskom, and S.~S. Ghosh, ``Nipype: a flexible, lightweight and extensible
  neuroimaging data processing framework in python,'' {\em Frontiers in
  neuroinformatics}, vol.~5, p.~13, 2011.

\bibitem{jenkinson2012fsl}
M.~Jenkinson, C.~F. Beckmann, T.~E. Behrens, M.~W. Woolrich, and S.~M. Smith,
  ``Fsl,'' {\em Neuroimage}, vol.~62, no.~2, pp.~782--790, 2012.

\bibitem{fischl2002van}
B.~Fischl, D.~Salat, E.~Busa, M.~Albert, M.~Dieterich, and C.~Haselgrove,
  ``Whole brain segmentation: automated labeling of neuroanatomical structures
  in the human brain,'' {\em Neuron}, vol.~33, no.~3, pp.~341--355, 2002.

\bibitem{avants2011reproducible}
B.~B. Avants, N.~J. Tustison, G.~Song, P.~A. Cook, A.~Klein, and J.~C. Gee, ``A
  reproducible evaluation of ants similarity metric performance in brain image
  registration,'' {\em Neuroimage}, vol.~54, no.~3, pp.~2033--2044, 2011.

\bibitem{penny2011statistical}
W.~D. Penny, K.~J. Friston, J.~T. Ashburner, S.~J. Kiebel, and T.~E. Nichols,
  {\em Statistical parametric mapping: the analysis of functional brain
  images}.
\newblock Elsevier, 2011.

\bibitem{morosan2001human}
P.~Morosan, J.~Rademacher, A.~Schleicher, K.~Amunts, T.~Schormann, and
  K.~Zilles, ``Human primary auditory cortex: cytoarchitectonic subdivisions
  and mapping into a spatial reference system,'' {\em Neuroimage}, vol.~13,
  no.~4, pp.~684--701, 2001.

\bibitem{moerel2014anatomical}
M.~Moerel, F.~De~Martino, and E.~Formisano, ``An anatomical and functional
  topography of human auditory cortical areas,'' {\em Frontiers in
  neuroscience}, vol.~8, p.~225, 2014.

\bibitem{besle2019human}
J.~Besle, O.~Mougin, R.-M. S{\'a}nchez-Panchuelo, C.~Lanting, P.~Gowland,
  R.~Bowtell, S.~Francis, and K.~Krumbholz, ``Is human auditory cortex
  organization compatible with the monkey model? contrary evidence from
  ultra-high-field functional and structural mri,'' {\em Cerebral Cortex},
  vol.~29, no.~1, pp.~410--428, 2019.

\bibitem{gulban2020improving}
O.~F. Gulban, R.~Goebel, M.~Moerel, D.~Zachlod, H.~Mohlberg, K.~Amunts, and
  F.~De~Martino, ``Improving a probabilistic cytoarchitectonic atlas of
  auditory cortex using a novel method for inter-individual alignment,'' {\em
  Elife}, vol.~9, p.~e56963, 2020.

\bibitem{benjamini1995controlling}
Y.~Benjamini and Y.~Hochberg, ``Controlling the false discovery rate: a
  practical and powerful approach to multiple testing,'' {\em Journal of the
  Royal statistical society: series B (Methodological)}, vol.~57, no.~1,
  pp.~289--300, 1995.

\bibitem{holm1979simple}
S.~Holm, ``A simple sequentially rejective multiple test procedure,'' {\em
  Scandinavian journal of statistics}, pp.~65--70, 1979.

\bibitem{rosa2010bayesian}
M.~J. Rosa, S.~Bestmann, L.~Harrison, and W.~Penny, ``Bayesian model selection
  maps for group studies,'' {\em Neuroimage}, vol.~49, no.~1, pp.~217--224,
  2010.

\bibitem{stephan2009bayesian}
K.~E. Stephan, W.~D. Penny, J.~Daunizeau, R.~J. Moran, and K.~J. Friston,
  ``Bayesian model selection for group studies,'' {\em Neuroimage}, vol.~46,
  no.~4, pp.~1004--1017, 2009.

\bibitem{klein2014stimulus}
C.~Klein, W.~von~der Behrens, and B.~H. Gaese, ``Stimulus-specific adaptation
  in field potentials and neuronal responses to frequency-modulated tones in
  the primary auditory cortex,'' {\em Brain topography}, vol.~27, no.~4,
  pp.~599--610, 2014.

\bibitem{thomas2012stimulus}
J.~M. Thomas, C.~Morse, L.~Kishline, A.~O’Brien-Lambert, A.~Simonton, K.~E.
  Miller, and E.~Covey, ``Stimulus-specific adaptation in specialized neurons
  in the inferior colliculus of the big brown bat, eptesicus fuscus,'' {\em
  Hearing research}, vol.~291, no.~1-2, pp.~34--40, 2012.

\bibitem{yue2007response}
Q.~Yue, J.~H. Casseday, and E.~Covey, ``Response properties and location of
  neurons selective for sinusoidal frequency modulations in the inferior
  colliculus of the big brown bat,'' {\em Journal of neurophysiology}, vol.~98,
  no.~3, pp.~1364--1373, 2007.

\bibitem{parras2021posterior}
G.~G. Parras, L.~Casado-Rom{\'a}n, E.~Schr{\"o}ger, and M.~S. Malmierca, ``The
  posterior auditory field is the chief generator of prediction error signals
  in the auditory cortex,'' {\em NeuroImage}, p.~118446, 2021.

\bibitem{glendenning1998comparative}
K.~Glendenning and R.~Masterton, ``Comparative morphometry of mammalian central
  auditory systems: variation in nuclei and form of the ascending system,''
  {\em Brain, behavior and evolution}, vol.~51, no.~2, pp.~59--89, 1998.

\bibitem{keller2018predictive}
G.~B. Keller and T.~D. Mrsic-Flogel, ``Predictive processing: a canonical
  cortical computation,'' {\em Neuron}, vol.~100, no.~2, pp.~424--435, 2018.

\bibitem{paavilainen2013mismatch}
P.~Paavilainen, ``The mismatch-negativity (mmn) component of the auditory
  event-related potential to violations of abstract regularities: a review,''
  {\em International journal of psychophysiology}, vol.~88, no.~2,
  pp.~109--123, 2013.

\bibitem{bendixen2009heard}
A.~Bendixen, E.~Schr{\"o}ger, and I.~Winkler, ``I heard that coming:
  event-related potential evidence for stimulus-driven prediction in the
  auditory system,'' {\em Journal of Neuroscience}, vol.~29, no.~26,
  pp.~8447--8451, 2009.

\bibitem{wacongne2011evidence}
C.~Wacongne, E.~Labyt, V.~van Wassenhove, T.~Bekinschtein, L.~Naccache, and
  S.~Dehaene, ``Evidence for a hierarchy of predictions and prediction errors
  in human cortex,'' {\em Proceedings of the National Academy of Sciences},
  vol.~108, no.~51, pp.~20754--20759, 2011.

\bibitem{chennu2016silent}
S.~Chennu, V.~Noreika, D.~Gueorguiev, Y.~Shtyrov, T.~A. Bekinschtein, and
  R.~Henson, ``Silent expectations: dynamic causal modeling of cortical
  prediction and attention to sounds that weren't,'' {\em Journal of
  Neuroscience}, vol.~36, no.~32, pp.~8305--8316, 2016.

\bibitem{blank2016prediction}
H.~Blank and M.~H. Davis, ``Prediction errors but not sharpened signals
  simulate multivoxel fmri patterns during speech perception,'' {\em PLoS
  biology}, vol.~14, no.~11, p.~e1002577, 2016.

\bibitem{ylinen2016predictive}
S.~Ylinen, M.~Huuskonen, K.~Mikkola, E.~Saure, T.~Sinkkonen, and
  P.~Paavilainen, ``Predictive coding of phonological rules in auditory cortex:
  A mismatch negativity study,'' {\em Brain and language}, vol.~162,
  pp.~72--80, 2016.

\bibitem{blank2018neural}
H.~Blank, M.~Spangenberg, and M.~H. Davis, ``Neural prediction errors
  distinguish perception and misperception of speech,'' {\em Journal of
  Neuroscience}, vol.~38, no.~27, pp.~6076--6089, 2018.

\bibitem{heilbron2021hierarchy}
M.~Heilbron, K.~Armeni, J.-M. Schoffelen, P.~Hagoort, and F.~P. de~Lange, ``A
  hierarchy of linguistic predictions during natural language comprehension,''
  {\em bioRxiv}, pp.~2020--12, 2021.

\bibitem{malmierca2015auditory}
M.~S. Malmierca, ``Auditory system,'' in {\em The rat nervous system},
  pp.~865--946, Elsevier, 2015.

\bibitem{anderson2013effect}
L.~Anderson and M.~Malmierca, ``The effect of auditory cortex deactivation on
  stimulus-specific adaptation in the inferior colliculus of the rat,'' {\em
  European Journal of Neuroscience}, vol.~37, no.~1, pp.~52--62, 2013.

\bibitem{bauerle2011stimulus}
P.~B{\"a}uerle, W.~von~der Behrens, M.~K{\"o}ssl, and B.~H. Gaese,
  ``Stimulus-specific adaptation in the gerbil primary auditory thalamus is the
  result of a fast frequency-specific habituation and is regulated by the
  corticofugal system,'' {\em Journal of Neuroscience}, vol.~31, no.~26,
  pp.~9708--9722, 2011.

\bibitem{plakas2013impaired}
A.~Plakas, T.~van Zuijen, T.~van Leeuwen, J.~M. Thomson, and A.~van~der Leij,
  ``Impaired non-speech auditory processing at a pre-reading age is a
  risk-factor for dyslexia but not a predictor: an erp study,'' {\em Cortex},
  vol.~49, no.~4, pp.~1034--1045, 2013.

\bibitem{neuhoff2012evidence}
N.~Neuhoff, J.~Bruder, J.~Bartling, A.~Warnke, H.~Remschmidt,
  B.~M{\"u}ller-Myhsok, and G.~Schulte-K{\"o}rne, ``Evidence for the late mmn
  as a neurophysiological endophenotype for dyslexia,'' {\em PloS one}, vol.~7,
  no.~5, p.~e34909, 2012.

\bibitem{gu2020auditory}
C.~Gu and H.-Y. Bi, ``Auditory processing deficit in individuals with dyslexia:
  A meta-analysis of mismatch negativity,'' {\em Neuroscience \& Biobehavioral
  Reviews}, 2020.

\bibitem{daliri2015electrophysiological}
A.~Daliri and L.~Max, ``Electrophysiological evidence for a general auditory
  prediction deficit in adults who stutter,'' {\em Brain and Language},
  vol.~150, pp.~37--44, 2015.

\bibitem{van2017predictive}
G.~I. van Schalkwyk, F.~R. Volkmar, and P.~R. Corlett, ``A predictive coding
  account of psychotic symptoms in autism spectrum disorder,'' {\em Journal of
  autism and developmental disorders}, vol.~47, no.~5, pp.~1323--1340, 2017.

\bibitem{fryer2020deficits}
S.~L. Fryer, B.~J. Roach, H.~K. Hamilton, P.~Bachman, A.~Belger, R.~E.
  Carri{\'o}n, E.~Duncan, J.~Johannesen, G.~A. Light, M.~Niznikiewicz, {\em
  et~al.}, ``Deficits in auditory predictive coding in individuals with the
  psychosis risk syndrome: Prediction of conversion to psychosis.,'' {\em
  Journal of Abnormal Psychology}, vol.~129, no.~6, p.~599, 2020.

\bibitem{perez2014automatic}
V.~B. Perez, S.~W. Woods, B.~J. Roach, J.~M. Ford, T.~H. McGlashan, V.~H.
  Srihari, and D.~H. Mathalon, ``Automatic auditory processing deficits in
  schizophrenia and clinical high-risk patients: forecasting psychosis risk
  with mismatch negativity,'' {\em Biological psychiatry}, vol.~75, no.~6,
  pp.~459--469, 2014.

\end{thebibliography}
\bibliographystyle{ieeetr}
 
\clearpage
\appendix
\renewcommand{\thefigure}{S\arabic{figure}}
\renewcommand{\thetable}{S\arabic{table}}
\setcounter{figure}{0}
\setcounter{table}{0}

\section*{Supplementary Material}

    \begin{figure}[h!]
        \centering
        \includegraphics[scale=0.4]{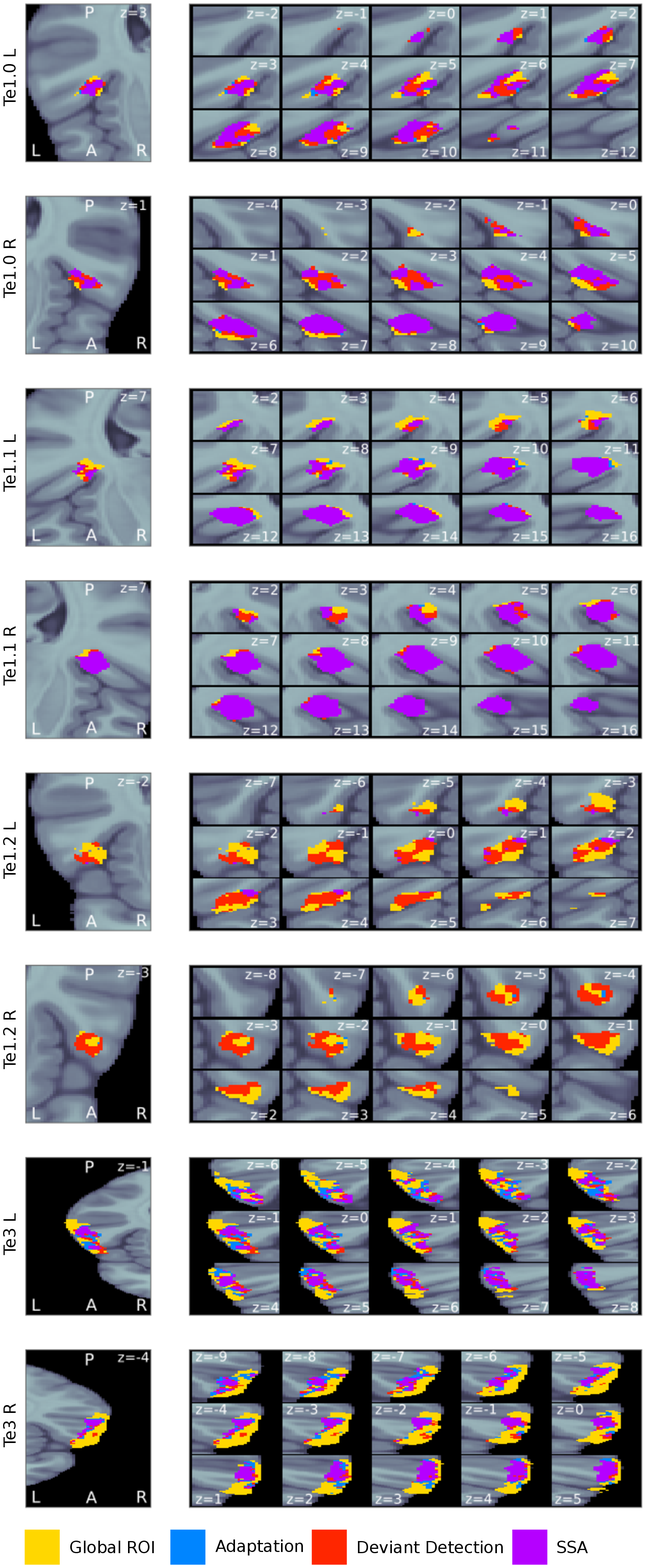}%{Figures/SSAROIsTotal.png}
        \caption[Pure Tones: Adaptation, Deviant Detection, and SSA Voxels]{\textbf{Adaptation, deviant detection, and SSA to pure tones within all eight anatomical ROIs.} Voxels exhibiting significant ($p < 0.05$, FDR-corrected) adaptation (blue indicates adaptation alone, purple indicates SSA which entails adaptation), deviant detection (red indicates deviant detection, purple indicates SSA which entails deviant detection), and SSA (purple) are shown for each anatomical ROI (yellow).}
        \label{fig:ssaROIS2D}
    \end{figure}
    \clearpage

    \begin{figure}[h!]
        \centering
        \includegraphics[scale=0.5]{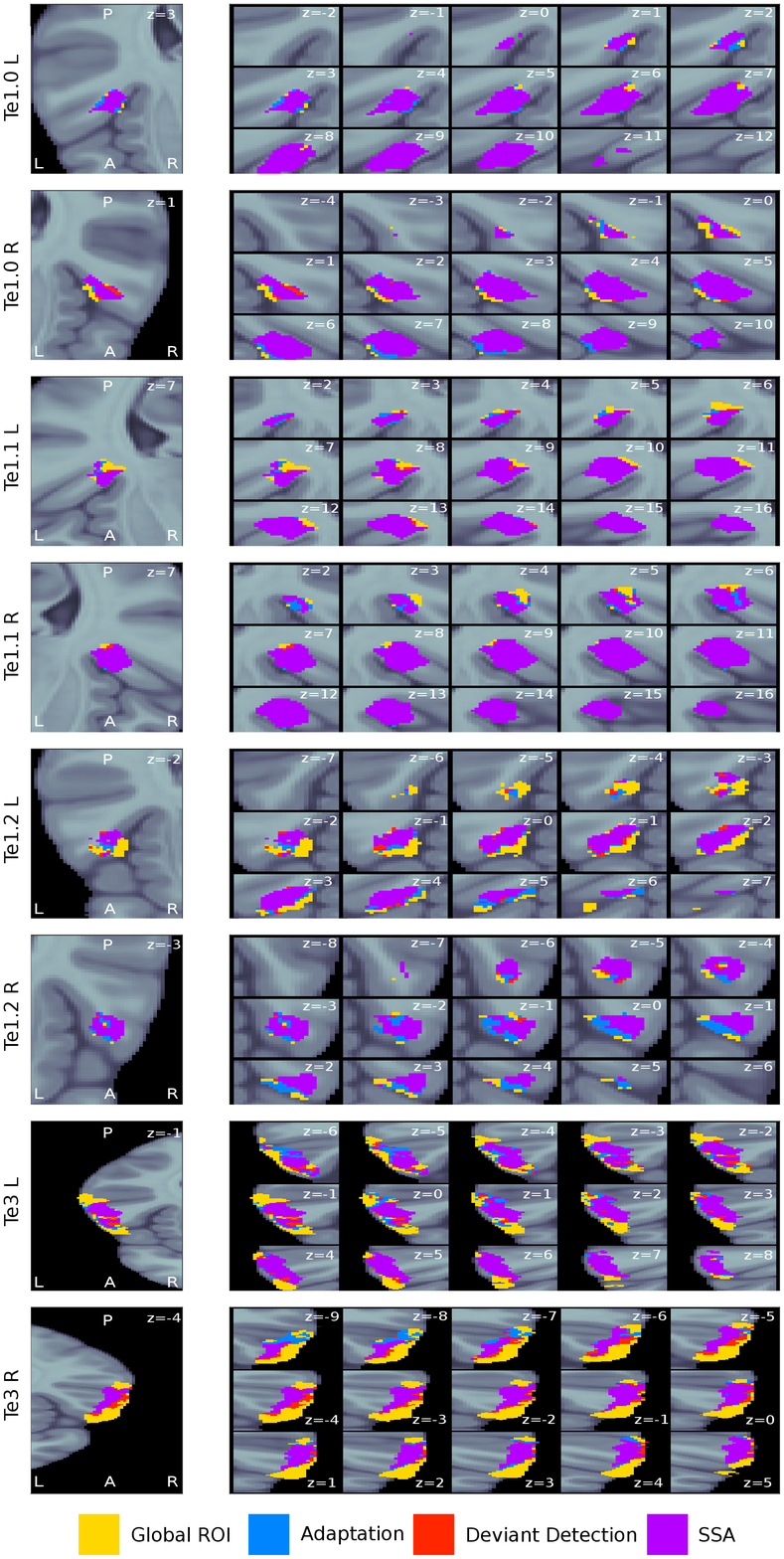}%{Figures/SSAROIsSweeps.png}
        \caption[FM-Sweeps: Adaptation, Deviant Detection, and SSA Voxels]{\textbf{Adaptation, deviant detection, and SSA to FM-sweeps within all eight anatomical ROIs.} Voxels exhibiting significant ($p < 0.05$, FDR-corrected) adaptation (blue indicates adaptation alone, purple indicates SSA which entails adaptation), deviant detection (red indicates deviant detection alone, purple indicates SSA which entails deviant detection), and SSA (purple) are shown for each anatomical ROI (yellow).}
        \label{fig:sweepsssaROIS2D}
    \end{figure}

    \begin{figure}[h!]
        \centering
        \includegraphics[scale=0.75]{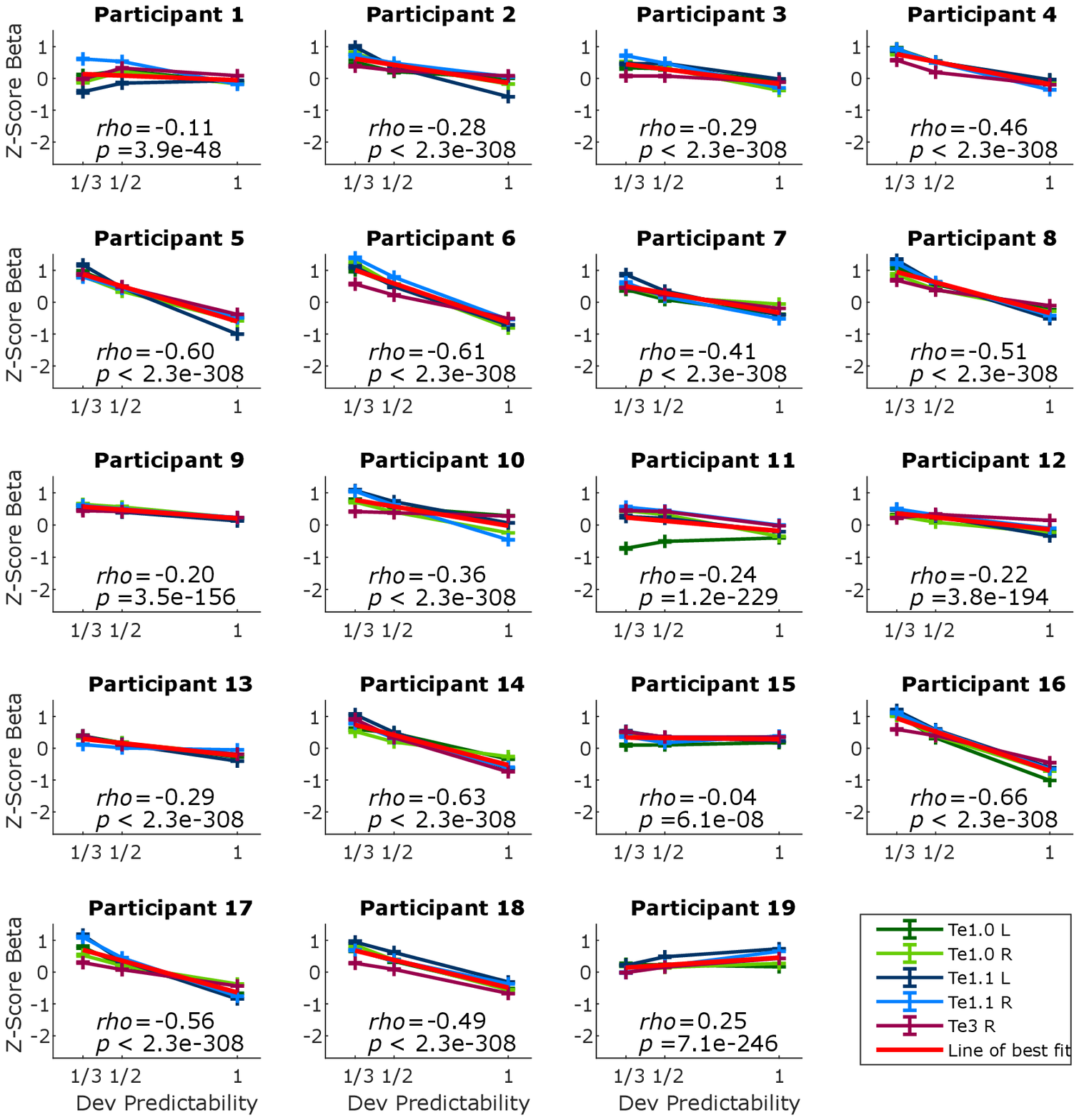}%{FiguresDraft2/Correlations_Subject_uncorrThick.png}
        \caption[Pure Tones: Participant-Level Correlational Analysis]{\textbf{Spearman's rank correlation between deviant predictability and standardized beta estimates for each participant of the pure tone experiment.} Deviant predictability is shown on the x-axis (1/3 for deviants in position four, 1/2 for deviants in position five, and 1 for deviants in position six). The y-axis shows the respective mean standardized beta estimates.}
        \label{fig:correlationssubPT}
        \end{figure}
    \clearpage

    \begin{figure}[ht!]
        \centering
        \includegraphics[scale=0.7]{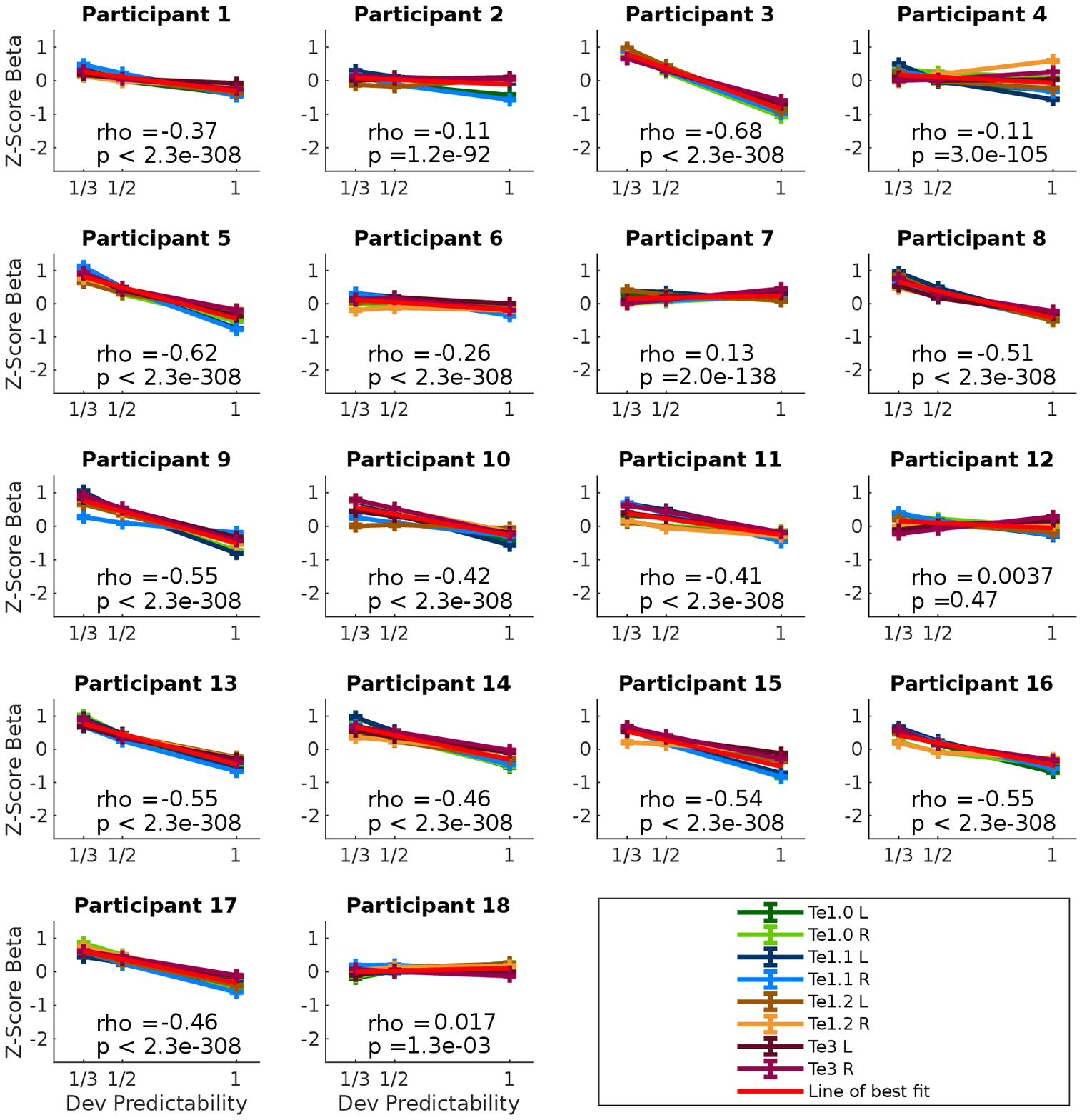}%{FiguresDraft2/Correlations_Subject_Sweeps_thick.png}
        \caption[FM-Sweeps: Participant-Level Correlational Analysis]{\textbf{Spearman's rank correlation between deviant predictability and standardized beta estimates for each participant of the FM-sweep experiment.} Deviant predictability is shown on the x-axis (1/3 for deviants in position four, 1/2 for deviants in position five, and 1 for deviants in position six). The y-axis shows the respective mean standardized beta estimates. The last panel shows the distribution of the correlation coefficient per SSA ROI.}
        \label{fig:correlationssubSweeps}
        \end{figure}
    \clearpage

    \begin{table}[h!]
        \centering
        \begin{tabular}{ c c c c c c }
            \hline
            Hypothesis       &    Te1.0 L   &    Te1.0 R   &   Te1.1 L    &    Te1.1 R    &     Te3 R     \\ 
            \hline
            $std0 > std2$    & $p = 0.0024$ & $p = 0.0029$ & $p = 0.0026$ & $p = 0.0025$  & $p = 0.0024$  \\
                             & $d = 1.93$   & $d = 2.66$   & $d = 2.78$   & $d = 2.68$    & $d = 2.77$    \\
            \hline
            $dev4 > std2$    & $p = 0.0047$ & $p = 0.0023$ &  $p = 0.0031$ & $p = 0.0021$ & $p = 0.0021$  \\
                             & $d = 2.19$   & $d = 3.51$   & $d = 3.17$    & $d = 3.93$   & $d = 3.20$    \\
            \hline
            $dev4 > dev6$    & $p = 0.0030$ & $p = 0.0028$ &  $p = 0.0037$ & $p = 0.0035$ & $p = 0.0037$  \\
                             & $d = 1.98$   & $d = 2.71$   & $d = 2.39$    & $d = 2.64$   & $d = 1.83$    \\
            \hline
            $dev4 > dev5$    & $p = 0.0063$ & $p = 0.0039$ &  $p = 0.0055$ & $p = 0.0034$ & $p = 0.0223$  \\
                             & $d = 0.67$   & $d = 1.14$   & $d = 0.98$    & $d = 1.09$   & $d = 0.74$    \\
            \hline
            $dev5 > dev6$    & $p = 0.0028$ & $p = 0.0027$ &  $p = 0.0027$ & $p = 0.0035$ & $p = 0.0032$  \\
                             & $d = 1.80$   & $d = 2.37$   & $d = 2.12$    & $d = 2.21$   & $d = 1.56$    \\
            \hline
            $dev6 \neq std2$ & $p = 0.6009$ & $p = 0.7961$ &  $p = 0.6820$ & $p = 0.0395$ & $p = 0.4662$  \\
                             & $d = 0.11$   & $d = 0.23$   & $d = 0.37$    & $d = 0.69$   & $d = 0.48$    \\
            \hline
        \end{tabular}
        \caption[Pure Tones: Group-Level Wilcoxon Sign Rank Tests]{\textbf{Statistics of the group-level Wilcoxon sign rank tests for the pure tone data.} The indicated hypotheses refer to the alternative hypotheses of the tests. The comparison of $dev6$ and $std2$ was conducted using two-tailed sign rank tests; all other contrasts were tested using one-sided sign rank tests. All $p$-values were corrected for 30 comparisons using the Holm-Bonferroni method. Effect size $d$: Cohen's $d$.}
        \label{tab:groupSignrankPT}
    \end{table}

\begin{table}
    \centering
    \begin{tabular}{ c c c c c}
        \hline
        Hypothesis       & Te1.0 L & Te1.0 R & Te1.1 L & Te1.1 R \\ [0.5ex]
        \hline
        $std0 > std2$    & $p = 0.0057$ & $p = 0.0057$ & $p = 0.0051$ & $p = 0.0050$ \\
                         & $d = 3.33$   & $d = 2.95$   & $d = 4.41$   & $d = 4.07$   \\
        \hline
        $dev4 > std2$    & $p = 0.0077$ & $p = 0.0104$ & $p = 0.0055$ & $p = 0.0049$ \\
                         & $d = 2.66$   & $d = 2.45$   & $d = 3.44$   & $d = 3.24$   \\
        \hline
        $dev4 > dev6$    & $p = 0.0060$ & $p = 0.0114$ & $p = 0.0056$ & $p = 0.0054$ \\
                         & $d = 2.91$   & $d = 2.26$   & $d = 3.46$   & $d = 3.29$   \\
        \hline
        $dev4 > dev5$    & $p = 0.0074$ & $p = 0.0100$ & $p = 0.0053$ & $p = 0.0059$ \\
                         & $d = 1.08$   & $d = 0.94$   & $d = 1.26$   & $d = 1.35$   \\
        \hline
        $dev5 > dev6$    & $p = 0.0056$ & $p = 0.0109$ & $p = 0.0052$ & $p = 0.0051$ \\
                         & $d = 2.51$   & $d = 1.91$   & $d = 3.02$   & $d = 2.84$   \\
        \hline 
        $dev6 \neq std2$ & $p = 0.3888$ & $p = 0.3720$ & $p = 0.6543$ & $p = 0.1057$ \\
                         & $d = -0.48$  & $d = -0.25$  & $d = -0.26$  & $d = -0.63$  \\
        \hline
        \\
        \hline
        Hypothesis       &  Te1.2 L & Te1.2 R & Te3 L & Te3 R \\ [0.5ex]
        \hline
        $std0 > std2$    & $p = 0.0107$ & $p = 0.0050$ & $p = 0.0053$ & $p = 0.0067$ \\
                         & $d = 2.12$   & $d = 2.35$   & $d = 2.76$   & $d = 2.65$   \\
        \hline
        $dev4 > std2$    & $p = 0.0169$ & $p = 0.0103$ & $p = 0.0096$ & $p = 0.0092$ \\
                         & $d = 2.07$   & $d = 1.86$   & $d = 2.44$   & $d = 2.52$   \\
        \hline
        $dev4 > dev6$    & $p = 0.0095$ & $p = 0.0509$ & $p = 0.0199$ & $p = 0.0159$ \\
                         & $d = 2.25$   & $d = 1.43$   & $d = 1.93$   & $d = 1.94$   \\
        \hline
        $dev4 > dev5$    & $p = 0.0104$ & $p = 0.0663$ & $p = 0.0292$ & $p = 0.0149$ \\
                         & $d = 0.83$   & $d = 0.48$   & $d = 0.66$   & $d = 0.68$   \\
        \hline
        $dev5 > dev6$    & $p = 0.0084$ & $p = 0.0319$ & $p = 0.0133$ & $p = 0.0161$ \\
                         & $d = 2.01$   & $d = 1.28$   & $d = 1.90$   & $d = 1.76$   \\
        \hline
        $dev6 \neq std2$ & $p = 0.3416$ & $p = 0.9183$ & $p = 0.0859$ & $p = 0.0765$ \\
                         & $d = -0.34$  & $d = 0.28$   & $d = 0.51$   & $d = 0.65$   \\
        \hline
    \end{tabular}
    \caption[FM-Sweeps: Group-Level Wilcoxon Sign Rank Tests]{\textbf{Statistics of the group-level Wilcoxon sign rank tests for the FM-sweep data.} The indicated hypotheses refer to the alternative hypotheses of the tests. The comparison of $dev6$ and $std2$ was conducted using two-tailed sign rank tests; all other contrasts were tested using one-sided sign rank tests. All $p$-values were corrected for 48 comparisons using the Holm-Bonferroni method. Effect size $d$: Cohen's $d$.}
    \label{tab:groupSignrankSweeps}
\end{table}
\clearpage

        \begin{figure}[h!]
            \centering
            \includegraphics[scale=0.4]{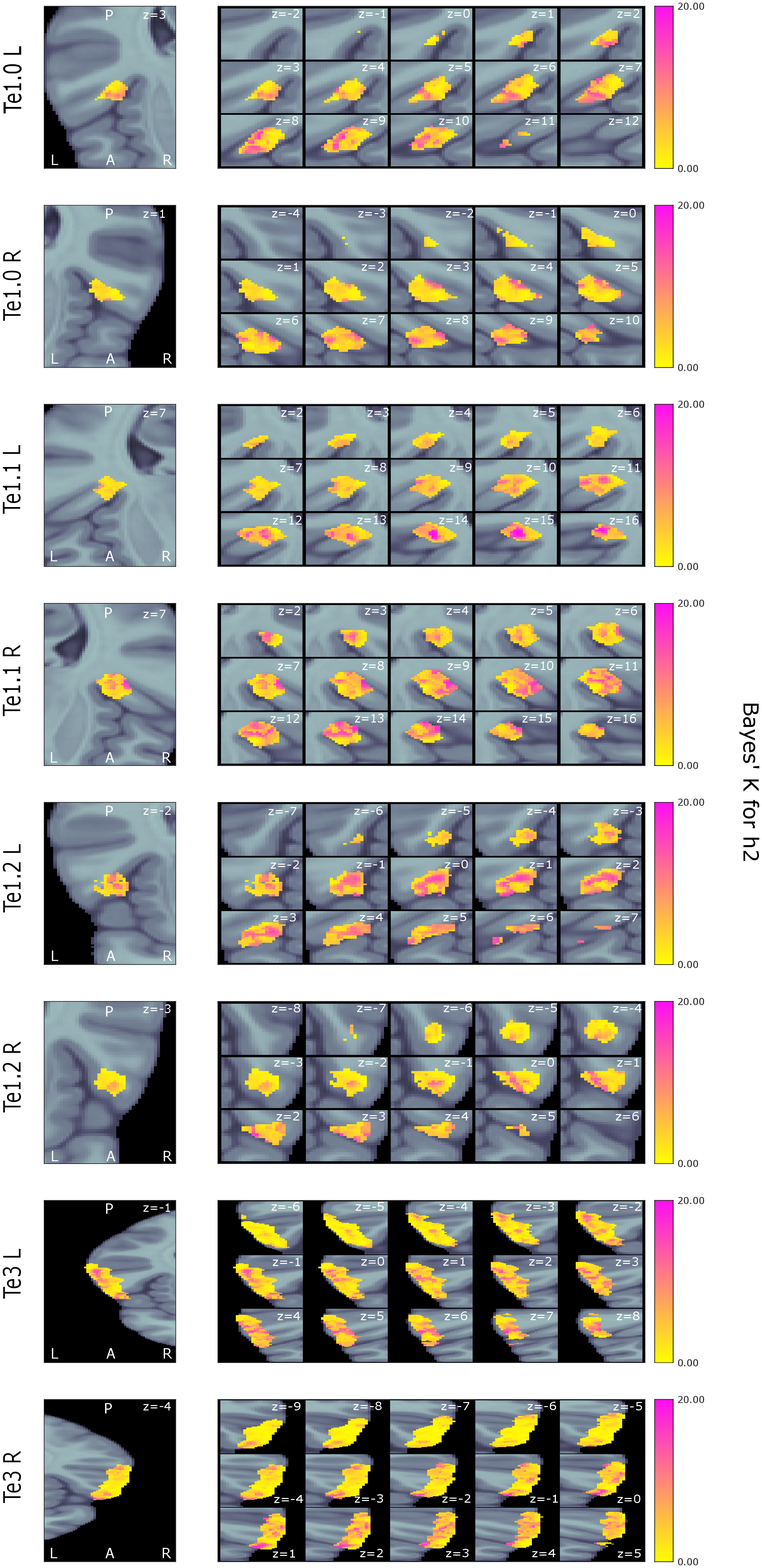}
            \caption[Bayesian Model Comparison: Habituation (h1) vs. prediction error (h2)]{Bayes' K map for h2; pure tones.}
            \label{fig:H2KPT}
        \end{figure}
        \clearpage

        \begin{figure}[h!]
            \centering
            \includegraphics[scale=0.4]{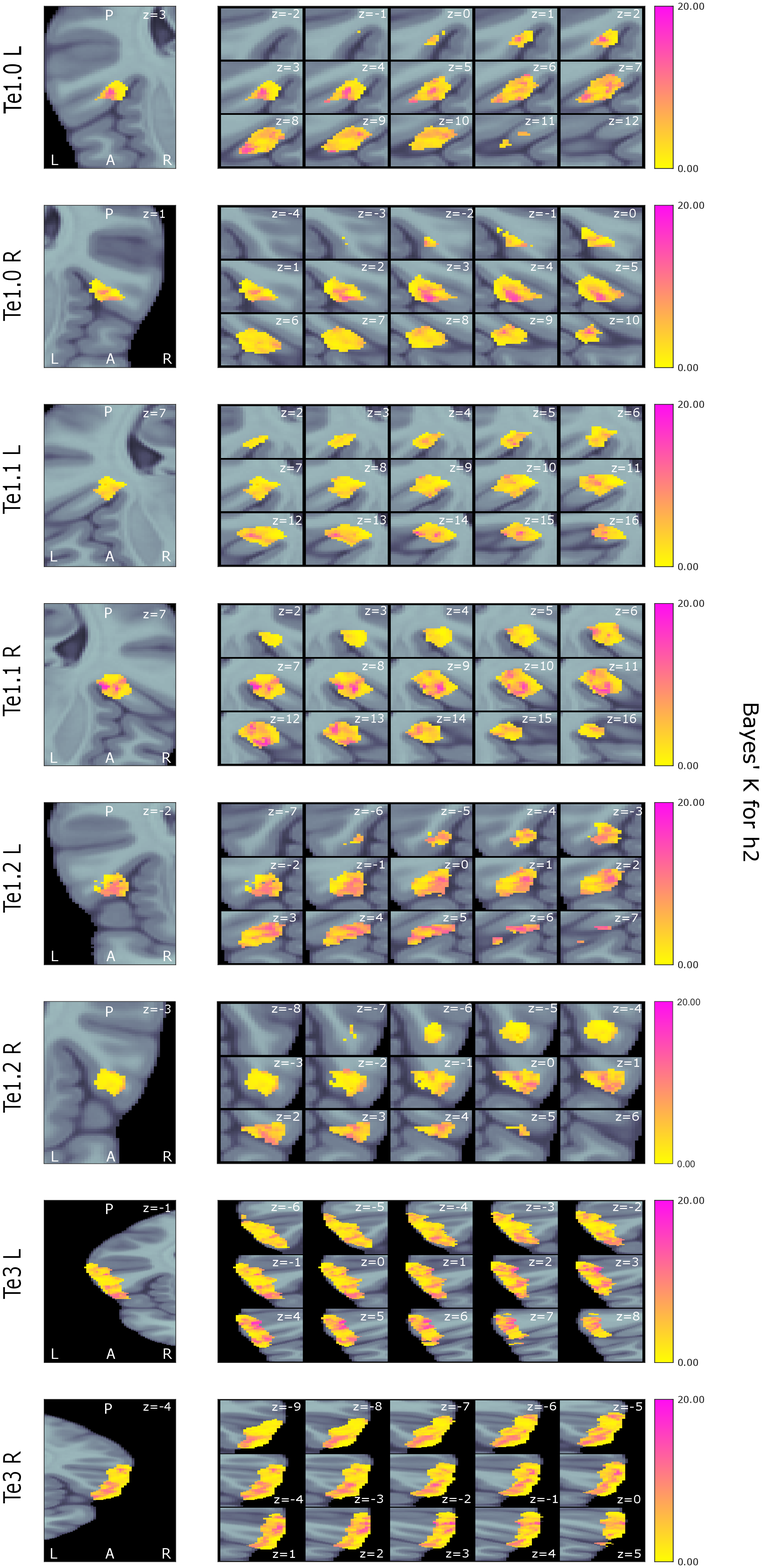}
            \caption[Bayesian Model Comparison: Habituation (h1) vs. prediction error (h2)]{Bayes' K map for h2, FM-sweeps}
            \label{fig:H2KSweeps}
        \end{figure}
        \clearpage

\end{document}